\documentclass[aps,prb,twocolumn,showpacs,showkeys,preprintnumbers,amsmath,amssymb,superscriptaddress]{revtex4-2}

\usepackage[utf8]{inputenc}
\usepackage{graphicx}
\usepackage{hyperref}
\usepackage{siunitx}
\usepackage{amsmath,amssymb}
\usepackage{textgreek}
\usepackage{multirow}
\usepackage{rotating}
\usepackage{color}
\usepackage{ulem}

\begin{document}
\title{Spin- and Angle-resolved Photoelectron Spectroscopy Study of the Quantum Spin Hall Insulator Bismuthene and its Precursor Phase}

\author{Niclas Tilgner}
\email{niclas.tilgner@physik.tu-chemnitz.de}
\affiliation{Institute of Physics, Chemnitz University of Technology, 09126 Chemnitz, Germany}
\affiliation{Research Center for Materials, Architectures and Integration of Nanomembranes (MAIN), 09126 Chemnitz, Germany}

\author{Andres David Pe\~{n}a Unigarro}
\affiliation{Institute of Physics, Chemnitz University of Technology, 09126 Chemnitz, Germany}

\author{Susanne Wolff}
\affiliation{Institute of Physics, Chemnitz University of Technology, 09126 Chemnitz, Germany}
\affiliation{Research Center for Materials, Architectures and Integration of Nanomembranes (MAIN), 09126 Chemnitz, Germany}

\author{Mats Leandersson}
\affiliation{MAX IV Laboratory, Lund University, 22484 Lund, Sweden}

\author{Craig Polley}
\affiliation{MAX IV Laboratory, Lund University, 22484 Lund, Sweden}

\author{Fabian G\"{o}hler}
\affiliation{Institute of Physics, Chemnitz University of Technology, 09126 Chemnitz, Germany}
\affiliation{Research Center for Materials, Architectures and Integration of Nanomembranes (MAIN), 09126 Chemnitz, Germany}

\author{Sibylle Gemming}
\affiliation{Institute of Physics, Chemnitz University of Technology, 09126 Chemnitz, Germany}

\author{Thomas Seyller}
\affiliation{Institute of Physics, Chemnitz University of Technology, 09126 Chemnitz, Germany}
\affiliation{Research Center for Materials, Architectures and Integration of Nanomembranes (MAIN), 09126 Chemnitz, Germany}

\author{Philip Sch\"{a}dlich}
\email{philip.schaedlich@physik.tu-chemnitz.de}
\affiliation{Institute of Physics, Chemnitz University of Technology, 09126 Chemnitz, Germany}
\affiliation{Research Center for Materials, Architectures and Integration of Nanomembranes (MAIN), 09126 Chemnitz, Germany}

\date{\today}

\begin{abstract}
    Recent studies have revealed that a confined bismuth layer at the graphene/SiC interface can be reversibly switched between a topologically trivial precursor phase and the quantum spin Hall insulator bismuthene. Here, we present a detailed spin- and angle-resolved photoelectron spectroscopy study of both structures, resolving the spin texture of their low-energy electronic states. Owing to the strong intrinsic spin-orbit coupling of bismuth and the asymmetric confinement potential at the interface, the valence bands of both structures are Rashba-split. We demonstrate the expected spin-momentum locking for both phases and Kramers' doublets through the investigation of the valence bands' spin polarization at multiple positions in reciprocal space.
\end{abstract}

{}

\maketitle

%=================
%	1 - Introduction 
%=================	

\section{Introduction}

    \noindent    
    The search for next-generation electronics has identified two-dimensional (2D) quantum spin Hall (QSH) materials at the forefront of condensed matter research. These materials, characterized by a bulk band gap and topologically protected spin-polarized edge states at zero magnetic field \cite{konig2007quantum, RothScience2009, KanePhys.Rev.Lett.2005, KanePhys.Rev.Lett.2005a}, offer a platform for dissipation-less charge transport and are pivotal for the development of quantum computing and spintronics \cite{Nayak_2008, hasan2010colloquium, TokuraNat.Phys.2017, han2018quantum, lodge2021atomically}.
    
    One material that has attracted considerable attention in this regard is bismuthene on SiC(0001), where a bulk band gap of approximately 0.8\,eV has been observed in experimental studies \cite{ReisScience2017}, suggesting the potential for the QSH effect to be achieved at room temperature. Crystallographically, bismuthene adopts a honeycomb lattice in which each Bi atom occupies a T$_1$ on-top site with respect to the underlying SiC substrate. More recently, it was demonstrated that Bi intercalation beneath epitaxial graphene on SiC(0001) gives rise to a topologically trivial precursor phase, which can be reversibly transformed into the topological insulator bismuthene via hydrogenation \cite{Tilgner2025, Gehrig2025}.
    
    Owing to the strong intrinsic spin-orbit coupling (SOC) of bismuth, together with the asymmetric confinement potential at the graphene/SiC interface, both the precursor and bismuthene phases are expected to exhibit spin-split valence bands mediated by the Rashba effect \cite{ReisScience2017, HsuNewJ.Phys.2015, li2018theoretical}. In this study, we use spin- and angle-resolved photoelectron spectroscopy (ARPES) to investigate the spin polarization of the low-energy electronic states of both phases, and compare the experimental results with density functional theory (DFT) calculations of the initial state spin textures.

    While previous spin-integrated ARPES measurements on bismuthene had already indicated the presence of an energy splitting in the valence band structure \cite{ReisScience2017, Tilgner2025, Gehrig2025}, analogous measurements of the precursor phase did not permit the observation of a band splitting \cite{Tilgner2025, SohnJ.KoreanPhys.Soc.2021}. Here, we exploit the enhanced resolving capability of spin-resolved ARPES \cite{dil2019spin} to unambiguously distinguish two oppositely spin-polarized contributions shifted in energy for both phases, thereby demonstrating the lifting of spin degeneracy. For both the precursor and bismuthene, the measured in-plane spin polarization is in good agreement with the theoretically predicted spin textures, evidencing Rashba-induced spin-momentum locking. However, in bismuthene we additionally observe a significant out-of-plane spin polarization, which cannot be explained by initial state arguments.

    The article is structured as follows: First, we briefly review the geometric and electronic structures of both Bi phases by presenting spin-integrated ARPES maps. Then, we proceed to show spin-resolved energy and momentum distribution curves (EDCs and MDCs), respectively, at distinct in-plane momenta and energies. The main part of the article is restricted to the presentation of spin polarizations and determined component intensities along these paths. The raw data is available in section VI of the Supplementary Information. A brief explanation of the measurement setup and principle can be found in the Methods section.

%=================
%	2 - Results & Discussion 
%=================	
    
\section{The precursor phase}

    \begin{figure*}[t!]
        \centering
        \includegraphics[scale=1]{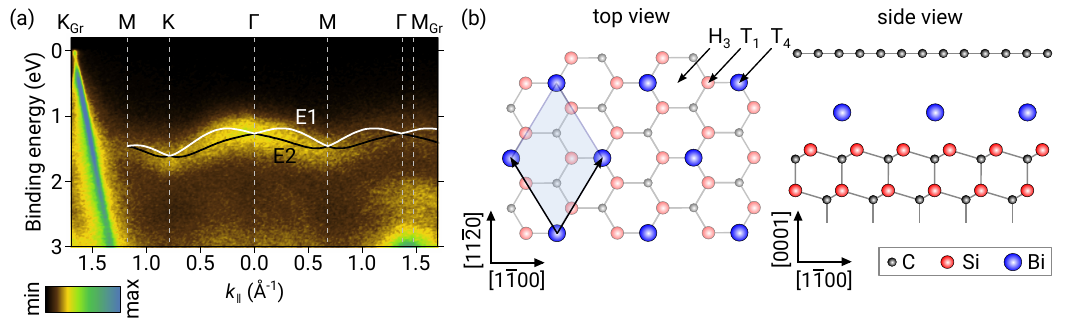}
        \caption{\textbf{Geometric and electronic structure of the precursor phase.}
        (a) Energy-momentum map acquired along the high-symmetry directions of graphene and the Bi lattice using a photon energy of $h\nu = 152.0$\,eV and $p$-polarization. The vertical dashed lines indicate the corresponding high-symmetry points. Points without an index correspond to the Bi lattice. Graphene's high-symmetry points are indicated by the index "Gr". Solid white and black curves overlaid on the experimental data represent the low-energy bands E1 and E2 obtained from DFT calculations for the Bi structure displayed in (b).
        (b) Ball-and-stick model of the precursor phase shown in top view (left) and side view (right). Bi atoms (blue) located at the graphene/SiC interface occupy T$_4$ hollow sites with a coverage of 1/3\,ML. In the left panel, the blue shaded area marks the surface unit cell of the Bi superstructure, which exhibits a $(\surd3 \times \surd3)R30^\circ$ symmetry relative to the substrate. For clarity, the graphene layer is omitted in the top view.
        } 
        \label{fig:intro_PRC}
    \end{figure*}

    \noindent
    We begin our discussion with the precursor phase. Fig. \ref{fig:intro_PRC}\,(a) presents a spin-integrated energy-momentum map recorded along the high-symmetry directions of graphene and the Bi lattice. A weakly dispersive feature associated with the precursor phase is distinctly visible in the vicinity of the \textGamma\ point. Note that the high-symmetry points of the bismuth lattices are labeled without an index throughout this article, while the equivalent points of graphene are denoted with the index "Gr".

    DFT calculations for Bi atoms adsorbed at T$_4$ hollow sites with a coverage of 1/3 monolayer (ML), depicted in the ball-and-stick model in Fig. \ref{fig:intro_PRC}\,(b), predict two Kramers-degenerate bands in the low-energy electronic structure. These calculated bands are overlaid in Fig. \ref{fig:intro_PRC}\,(a) as solid white (E1) and black (E2) lines and reproduce the experimentally observed dispersion well. Due to the comparatively large linewidth of the spin-integrated experimental spectra, the associated energy splitting ($\le 0.21$\,eV) cannot be directly resolved. Nevertheless, as demonstrated below, spin-resolved photoemission measurements reveal two oppositely polarized contributions to the photocurrent separated by a small energy offset, constituting direct evidence for the SOC-induced lifting of spin degeneracy.

    The identification of a Bi coverage of 1/3\,ML in the precursor phase also provides important insight into the structural transition toward bismuthene. In our earlier work (see Ref. \cite{Tilgner2025}), the precursor phase was assumed to possess a coverage of 2/3\,ML, corresponding to that of bismuthene (see below, in particular Fig. \ref{fig:intro_BME}\,(b), and Refs. \cite{Tilgner2025, Gehrig2025, ReisScience2017}). Within that framework, the phase transition, induced by hydrogenation, could be interpreted as a purely lateral displacement of the Bi layer. The present findings, however, demonstrate that the precursor contains only half of the previously assumed Bi density. Consequently, the conversion from the precursor phase to bismuthene must involve a lateral contraction. The area initially covered by the precursor phase therefore separates into regions of bismuthene and regions consisting exclusively of hydrogen-intercalated quasi-freestanding graphene (H/QFG). This interpretation is consistent with a recent scanning tunneling microscopy study \cite{Ngo2025quantum}, which reported bismuthene islands embedded within an H/QFG matrix and thereby supports the picture of a Bi layer contraction induced by hydrogenation.

    Within the revised structural model of the precursor phase proposed here, the Bi adatoms exhibit $sp^3$-hybridization. Three valence electrons participate in bonds that saturate the Si dangling bonds of the substrate, while the remaining fully occupied hybrid orbital points perpendicular to the surface and is expected to generate the observed surface state. The out-of-plane character is confirmed by the orbital projection of the Bi $p$-orbitals onto the calculated band structure, which demonstrates that the discussed surface state is solely of $p_z$-type (see section I of the Supplementary Information, specifically Fig. S1\,(b)). A comparable interpretation was previously suggested by Sohn \textit{et al}. \cite{SohnJ.KoreanPhys.Soc.2021}; however, they proposed adsorption at the H$_3$ site, whereas our earlier experiments unambiguously demonstrated occupation of the T$_4$ hollow site \cite{Tilgner2025, Tilgner2025_2}. \\
    
    \begin{figure*}[ht!]
        \centering
        \includegraphics[scale=1]{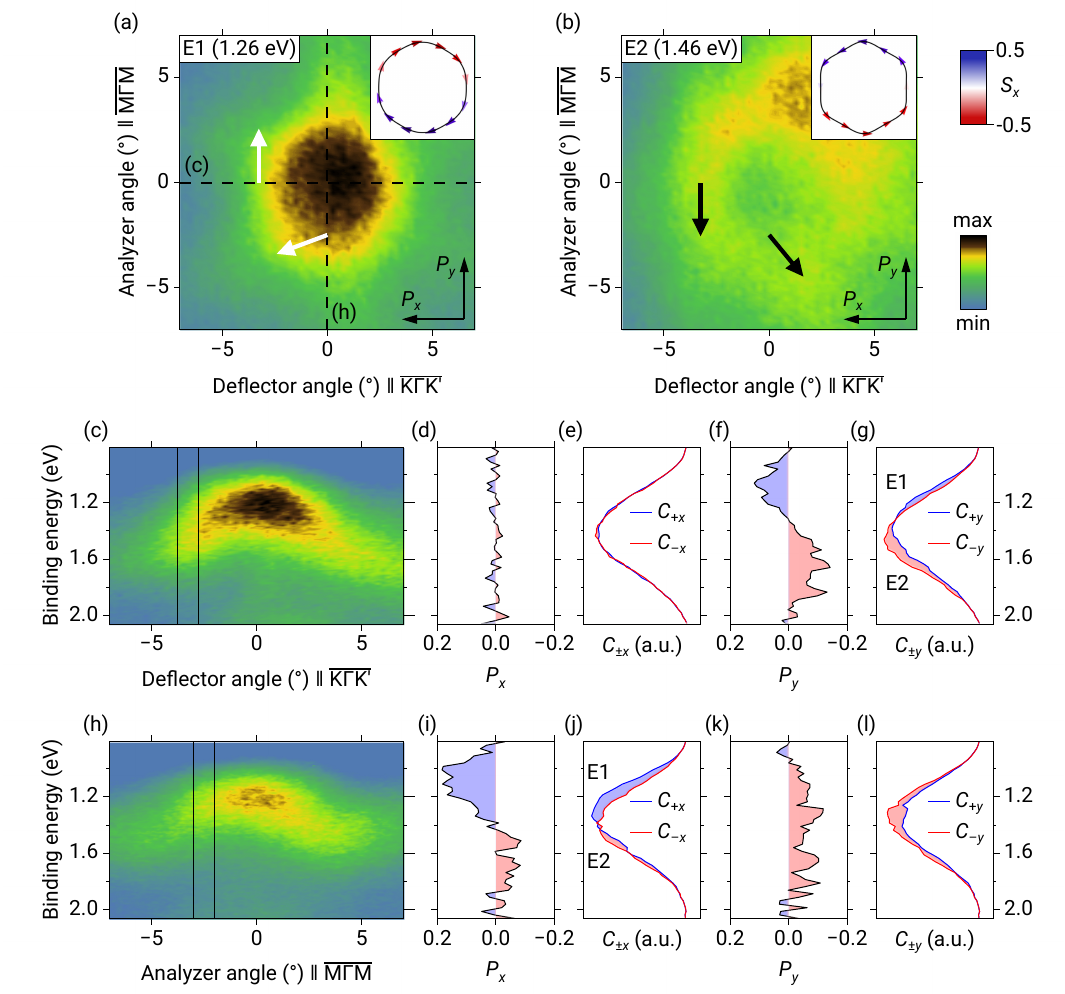}
        \caption{\textbf{Spin polarization of the precursor phase.}
        (a,b) Constant-energy maps extracted at binding energies of 1.26\,eV (a) and 1.46\,eV (b). White and black arrows superimposed on the maps indicate the experimentally determined spin polarization associated with the E1 and E2 states, respectively. The origin of each arrow corresponds to the angular position at which the spin-resolved EDCs were recorded, while the arrow orientation represents the measured spin polarization within the coordinate system shown in the lower right corner. Black dashed lines mark the positions of the energy-dispersive cuts presented in (c) and (h).
        The insets in (a) and (b) display the calculated spin textures of bands E1 and E2 at the respective binding energies of the constant-energy maps. Arrow orientations indicate the in-plane spin direction, whereas the color scale represents the projection onto the $S_x$ component. Blue (red) denotes positive (negative) values.
        (c,h) Energy-angle maps measured along the $\overline{\text{K}\Gamma\text{K'}}$ direction (c) and the $\overline{\text{M}\Gamma\text{M}}$ direction (h). The vertical black lines indicate the regions from which the EDCs were extracted with an angular integration width of $1^\circ$.
        (d-g) and (i-l) Corresponding in-plane spin polarizations $P_i$ together with the derived spin-resolved component intensities $C_{\pm i}$ with $i \in \{x,y\}$. Blue (red) represents positive (negative) spin polarization.
        All measurements were performed using a photon energy of $h\nu = 152.0$\,eV and $p$-polarization.
        } 
        \label{fig:spin_PRC}
    \end{figure*}

    Our investigation of the precursor phase is further extended by spin-resolved ARPES measurements. Spin-resolved EDCs were acquired at momentum positions where DFT predicts the largest energy splitting between the E1 and E2 states, namely between $\Gamma$ and the M or K points.

    As a reference, a three-dimensional (3D) spin-integrated dataset of the precursor-related electronic states was recorded. Two constant-energy slices extracted from this dataset are presented in Figs. \ref{fig:spin_PRC}\,(a) and (b). The selected binding energies correspond to the energies at which the spin-resolved components associated with E1 and E2 exhibit maximum intensity (see below).

    An energy-angle map extracted from the same 3D dataset along the horizontal black dashed line indicated in Fig. \ref{fig:spin_PRC}\,(a) is shown in (c). This cut corresponds to the $\overline{\text{K}\Gamma\text{K'}}$ direction of the precursor phase. A spin-resolved EDC was measured at the momentum position marked by the vertical black lines. The resulting in-plane spin polarization of the photoelectron beam along the $x$-direction, that is, "$P_x$" (see Methods section for details on determining the polarization), which is parallel to $\overline{\text{K}\Gamma\text{K'}}$ (see coordinate system in lower right corner of Figs. \ref{fig:spin_PRC}\,(a) and (b)), is displayed in Fig. \ref{fig:spin_PRC}\,(d). No measurable spin polarization is detected along $x$. Accordingly, decomposition of the EDC into components $C_{\pm x}$ corresponding to $P_x > 0$ ($C_{+x}$) and $P_x < 0$ ($C_{-x}$) contributions, shown in Fig. \ref{fig:spin_PRC}\,(e), does not reveal any distinction between the two.

    A different behavior emerges for the perpendicular in-plane spin component along the $y$-direction, i.e.\ parallel to $\overline{\text{M}\Gamma\text{M}}$. As shown in Fig. \ref{fig:spin_PRC}\,(f), the spin polarization is positive at lower binding energies and reverses sign toward higher binding energies. This observation directly demonstrates the lifting of spin degeneracy into two oppositely polarized states separated in energy. The corresponding decomposition of the EDC into $C_{+y}$ (blue) and $C_{-y}$ (red) contributions is presented in Fig. \ref{fig:spin_PRC}\,(g). An energy offset of approximately 0.2\,eV between the two components is observed, in excellent agreement with the splitting predicted by DFT (see Fig. \ref{fig:intro_PRC}\,(a)). The $C_{+y}$ contribution is therefore assigned to E1, whereas the $C_{-y}$ contribution is attributed to E2.

    To further clarify the origin of the precursor spin texture, the momentum position was varied in order to examine the evolution of the spin polarization. Fig. \ref{fig:spin_PRC}\,(h) displays an energy-angle map measured along the direction parallel to $\overline{\text{M}\Gamma\text{M}}$, corresponding to the vertical black dashed line indicated in Fig. \ref{fig:spin_PRC}\,(a). The associated $P_x$ polarization of the EDC, obtained in the region marked by the vertical black lines in Fig. \ref{fig:spin_PRC}\,(h), is shown in (i). In contrast to the previously discussed geometry, a finite spin polarization along $x$ is now clearly observed. The polarization is positive at lower binding energies and negative at higher binding energies. Consequently, the spin-resolved EDC shown in Fig. \ref{fig:spin_PRC}\,(j) exhibits an energy separation between the positive and negative components analogous to the splitting observed in (g). These results demonstrate that both E1 and E2 acquire a finite $P_x$ component upon moving along the constant-energy contours.

    The corresponding measurements for the $P_y$ polarization are presented in Figs. \ref{fig:spin_PRC}\,(k) and (l). Here, a nearly constant negative offset in $P_y$ is observed, independent of whether the signal originates from E1 or E2. This observation contrasts with the expectation of two oppositely polarized states. The origin of this behavior remains unclear.

    To summarize the experimentally determined spin texture of the precursor phase, arrows indicating the measured spin polarization were superimposed onto Figs. \ref{fig:spin_PRC}\,(a) and (b) for E1 and E2, respectively. The base of each arrow marks the momentum position at which the corresponding EDC was recorded. The arrow orientations were determined from the $P_x$ and $P_y$ values measured at the energies corresponding to the intensity maxima of the E1 and E2 components in Figs. \ref{fig:spin_PRC}\,(g) and (j). The resulting spin texture follows the constant-energy contours in an almost tangential manner, indicating Rashba-type spin-momentum locking.

    These experimental findings are consistent with theoretical calculations of the precursor phase spin texture, shown for E1 and E2 in the insets of Figs. \ref{fig:spin_PRC}\,(a) and (b), respectively. Both the tangential orientation and the winding direction of the spins are reproduced by the calculations.

    Finally, no measurable out-of-plane spin polarization was detected for the precursor phase (see Supplementary Information section II), suggesting that the observed spin texture originates predominantly from a Rashba-type contribution.

\section{The QSH phase}

    \begin{figure*}[ht!]
        \centering
        \includegraphics[scale=1]{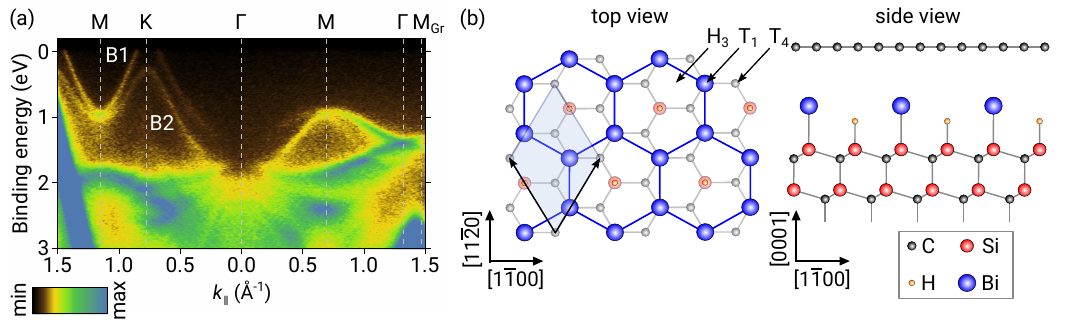}
        \caption{\textbf{Geometric and electronic structure of the bismuthene phase.}
        (a) Energy-momentum map acquired along the high-symmetry directions of graphene and bismuthene using a photon energy of $h\nu = 114.0$\,eV and $p$-polarization. The vertical dashed lines indicate the corresponding high-symmetry points. Points without an index correspond to the Bi lattice. Graphene's high-symmetry points are indicated by the index "Gr".
        (b) Ball-and-stick model of bismuthene shown in top view (left) and side view (right). Bi atoms (blue) located at the graphene/SiC interface occupy T$_1$ on-top sites and arrange in a honeycomb lattice with a coverage of 2/3\,ML. The terminating Si atom in the middle of the honeycomb is saturated by hydrogen (orange). In the left panel, the blue shaded area marks the surface unit cell of the Bi superstructure, which exhibits a $(\surd3 \times \surd3)R30^\circ$ symmetry relative to the substrate. For clarity, the graphene layer is omitted in the top view.
        } 
        \label{fig:intro_BME}
    \end{figure*}

    \noindent
    Upon transformation into bismuthene -- that is, upon hydrogenation of the precursor -- the electronic structure changes significantly, as seen in the spin-integrated energy-momentum map in Fig. \ref{fig:intro_BME}\,(a) and Refs. \cite{Tilgner2025, Gehrig2025}. The weakly dispersing precursor-related bands disappear while sharp Dirac-like bands now dominate the low-energy spectrum. Specifically, the two Rashba-split bands, labeled B1 and B2, are a hallmark of bismuthene formation -- the crystallographic structure of which is depicted schematically in Fig. \ref{fig:intro_BME}\,(b). As previously mentioned (see above and Refs. \cite{Tilgner2025, Gehrig2025}), partial hydrogenation of the substrate interface stabilizes the Bi atoms in a honeycomb pattern at T$_1$ on-top positions. This configuration establishes the orbital composition around the Fermi level necessary for forming the observed band structure \cite{ReisScience2017, HsuNewJ.Phys.2015, ZhouProc.Natl.Acad.Sci.2014, Zhou_2014}. Importantly, the remaining Bi-Si bond moves the $p_z$-contribution away from the Fermi level, which, consequently, is observed at higher binding energies, specifically between 1.5\,eV and 2.0\,eV in Fig. \ref{fig:intro_BME}\,(a). In the following, we focus on resolving the spin polarization of the B1 and B2 valence bands. \\

    \begin{figure*}[ht!]
        \centering
        \includegraphics[scale=1]{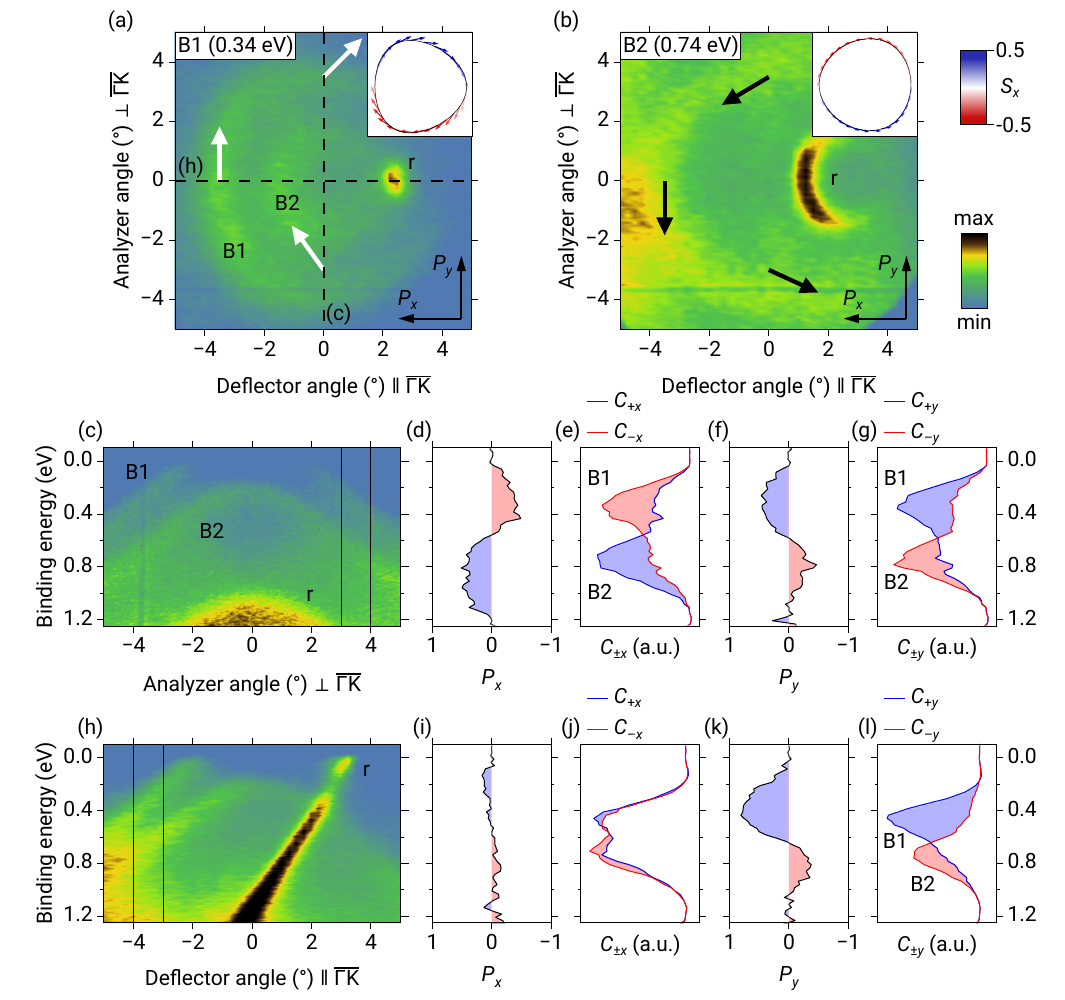}
        \caption{\textbf{In-plane spin polarization of the bismuthene phase.}
        (a,b) Constant-energy maps extracted at binding energies of 0.34\,eV (a) and 0.74\,eV (b). White and black arrows superimposed on the maps indicate the experimentally measured spin polarization associated with the B1 and B2 states, respectively. The base of each arrow marks the angular position at which the spin-resolved EDCs were recorded, while the arrow orientation reflects the spin polarization within the coordinate system shown in the lower right corner. The black dashed lines indicate the locations of the energy-angle cuts presented in (c) and (h).
        The insets in (a) and (b) display the calculated spin textures of bands B1 and B2 at the respective binding energies of the constant-energy maps. Arrow orientations indicate the in-plane spin direction, whereas the color scale represents the projection onto the $S_x$ component. Blue (red) denotes positive (negative) values.
        (c,h) Energy-angle maps measured perpendicular (c) and parallel (h) to the $\overline{\Gamma\text{K}}$ direction. Vertical black lines denote the momentum regions from which the EDCs were extracted using an angular integration width of $1^\circ$.
        (d-g) and (i-l) Corresponding in-plane spin polarizations $P_i$ together with the derived spin-resolved component intensities $C_{\pm i}$ with $i \in \{x,y\}$. Blue (red) denotes positive (negative) spin polarization.
        The measurements were performed with a photon energy of $h\nu = 39.0$\,eV and $p$-polarization.
        } 
        \label{fig:spin_BME}
    \end{figure*}

    First, a 3D spin-integrated ARPES dataset centered around the K point was acquired as a reference. The two constant-energy cuts extracted from this dataset which are shown in Figs. \ref{fig:spin_BME}\,(a) and (b) correspond to the maxima of the spin-resolved components associated with bands B1 and B2 (see below). In Fig. \ref{fig:spin_BME}\,(a), the contours of both bands are visible and labeled accordingly. At the larger binding energy used in Fig. \ref{fig:spin_BME}\,(b), only the contour of B2 remains visible due to the finite angular range. In both constant-energy maps, an additional sharp band, labeled "r", is visible, which is a replica of the graphene Dirac cone. The appearance of the replica depends on photon energy. However, this contribution does not affect the following analysis of the bismuthene-related bands.

    Fig. \ref{fig:spin_BME}\,(c) presents an energy-angle map extracted perpendicular to the $\overline{\Gamma\text{K}}$ direction, as indicated by the vertical dashed line in (a). The vertical black lines mark the angular region from which a spin-resolved EDC was obtained. The corresponding spin polarizations along the $x$- and $y$-directions, which are parallel and perpendicular to $\overline{\Gamma\text{K}}$, are shown in Figs. \ref{fig:spin_BME}\,(d) and (f), respectively. In both in-plane directions, a finite polarization is observed that reverses sign as a function of binding energy. Accordingly, the EDC can be decomposed into two oppositely in-plane polarized contributions: ($C_{-x}, C_{+y}$) at lower binding energies and ($C_{+x}, C_{-y}$) at higher binding energies, as illustrated in Figs. \ref{fig:spin_BME}\,(e) and (g). The maxima of these spin-resolved components coincide well with the energies at which bands B1 and B2 are observed in Fig. \ref{fig:spin_BME}\,(c). We therefore assign these contributions to the spin-split valence bands of bismuthene.

    A second spin-resolved EDC was recorded along the $\overline{\Gamma\text{K}}$ direction at the momentum position indicated by the vertical black lines in the energy-angle map shown in Fig. \ref{fig:spin_BME}\,(h). As demonstrated in Figs. \ref{fig:spin_BME}\,(i) and (j), no significant polarization along the $x$-direction is detected at this location. In contrast to the previously discussed geometry, the in-plane spin polarization of the bismuthene valence bands is here entirely dominated by the $y$-component, as shown in Figs. \ref{fig:spin_BME}\,(k) and (l).

    The experimentally determined spin polarizations for B1 and B2 are summarized by the white and black arrows superimposed in Figs. \ref{fig:spin_BME}\,(a) and (b), respectively. Along the $\overline{\Gamma\text{K}}$ direction (left arrows), the spin polarization follows the contours of the bands tangentially. For measurements performed along the perpendicular direction (top and bottom arrows), however, a clear deviation from a purely tangential orientation is observed. The measurement revealing the direction of the bottom arrows is presented in section III of the Supplementary Information, specifically in Fig. S3.

    In addition to the tangential component along $x$, both bands exhibit a pronounced polarization along the $y$-direction on either side of the K point, with opposite signs for B1 and B2. For each band, the sign of $P_y$ remains unchanged along the left part of the contour. Consequently, the $y$-component of the spin polarization points toward the K point on one side and away from it on the opposite side. This experimental observation is in excellent agreement with DFT calculations of the in-plane spin texture of bismuthene, as shown for bands B1 and B2 in the insets of Figs. \ref{fig:spin_BME}\,(a) and (b), respectively. The deviation from a purely tangential Rashba-like spin texture, which arises from terms linear in $k$, can be attributed to higher-order contributions to the SOC Hamiltonian. In particular, cubic terms allowed by the $C_{3v}$ symmetry of bismuthene become increasingly relevant at larger momenta relative to the K point, thereby giving rise to the observed non-tangential spin texture \cite{vajna2012higher, zhao2020purely}. Along the $\overline{\Gamma\text{K}}$ direction -- which, together with the sample normal, spans a mirror plane of the system -- non-tangential terms are not permitted by symmetry \cite{yaji2017spin}. \\

    \begin{figure*}[ht!]
        \centering
        \includegraphics[scale=1]{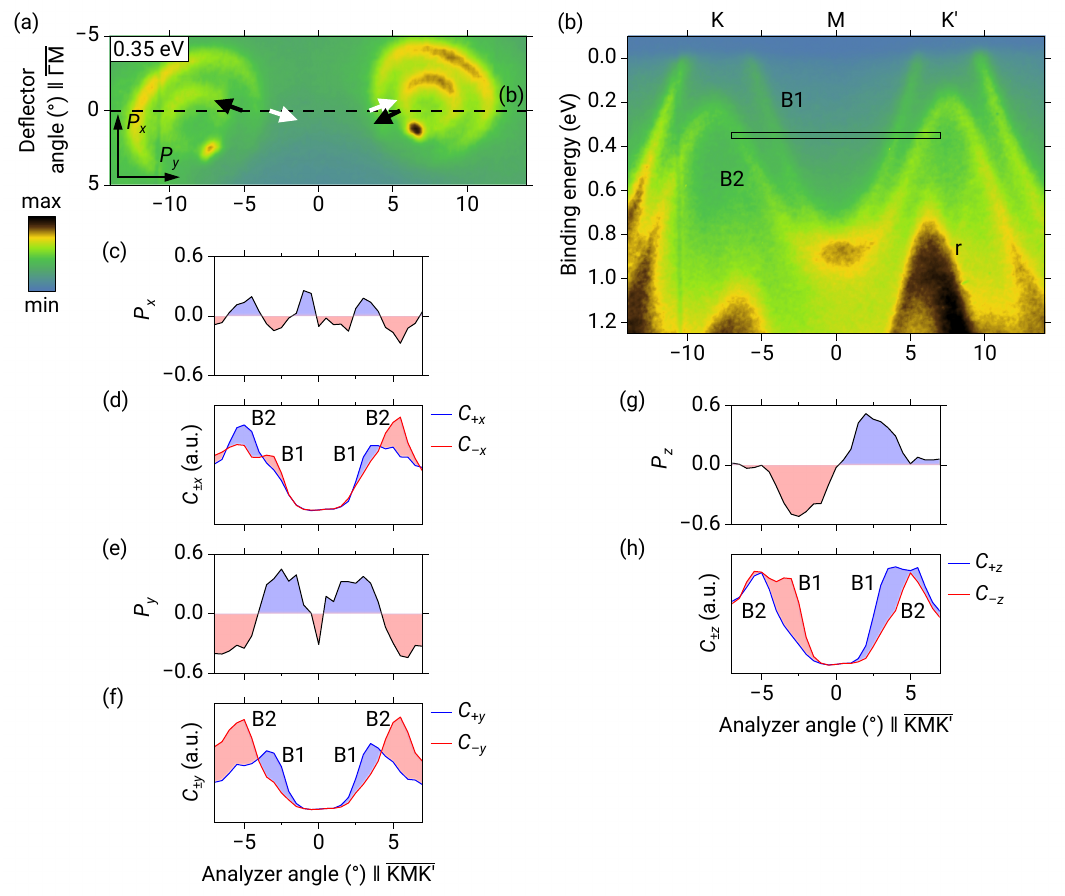}
        \caption{\textbf{Spin polarization of the bismuthene phase along $\overline{\text{KMK'}}$.}
        (a) Constant-energy map extracted at a binding energy of 0.35\,eV. White and black arrows superimposed on the map indicate the experimentally determined spin polarization associated with states B1 and B2, respectively. The origin of each arrow marks the angular position at which the maxima of the spin-resolved MDC components were observed, while the arrow orientation represents the spin polarization within the coordinate system shown in the lower left corner. The black dashed line indicates the location of the energy-angle cut presented in (b).
        (b) Energy-angle map measured along the $\overline{\text{KMK'}}$ direction. The black rectangle marks the energy interval from which the MDC was extracted using an energy acceptance window of 30\,meV.
        (c-f) Corresponding in-plane spin polarizations $P_i$ together with the derived spin-resolved component intensities $C_{\pm i}$ with $i \in \{x,y\}$.
        (g,h) Corresponding out-of-plane spin polarizations $P_z$ and the associated spin-resolved component intensities $C_{\pm z}$. Blue (red) denotes positive (negative) spin polarization.
        All measurements were performed at a photon energy of $h\nu = 39.0$\,eV and with $p$-polarization.
        } 
        \label{fig:spin_BME_KMK}
    \end{figure*}

    The investigation of the spin polarization of the bismuthene valence bands was further extended by performing a spin-resolved MDC measurement along the $\overline{\text{KMK'}}$ direction. A constant-energy map recorded at the binding energy used for the MDC extraction is shown in Fig. \ref{fig:spin_BME_KMK}\,(a). At this energy, all four hole pockets associated with bands B1 and B2 at the K and K' points are clearly resolved. The corresponding energy-angle map measured along the horizontal dashed line indicated in Fig. \ref{fig:spin_BME_KMK}\,(a) is presented in (b). The MDC was extracted from the region marked by the black rectangle, covering the B1 and B2 bands around both K and K'.

    The resulting in-plane spin polarizations and corresponding spin-resolved component intensities are displayed in Figs. \ref{fig:spin_BME_KMK}\,(c-f). From the maxima of the spin-resolved components, the in-plane spin polarization vectors were determined and visualized in Fig. \ref{fig:spin_BME_KMK}\,(a) as white and black arrows for B1 and B2, respectively. The experimentally observed spin texture is fully consistent with the results obtained previously at a single K point as well as with the theoretically predicted in-plane spin texture of bismuthene (compare Fig. \ref{fig:spin_BME}). In particular, both the winding direction and the non-tangential spin polarization are reproduced.

    Furthermore, these measurements demonstrate that the electronic states around K and K' exhibit the same spin winding direction. This behavior is, however, an intrinsic consequence of the sample system under investigation. As discussed in greater detail by Refs. \cite{Tilgner2025, Tilgner2025_2}, bismuthene forms on both S2 and S2* terraces of the 4H-SiC substrate, which are rotated by $60^{\circ}$ with respect to one another. As a result, photoemission measurements averaging over both rotational domains superimpose the S2 contribution from K with the S2* contribution from K', and vice versa. Since S2 and S2* are distributed equally across the surface, and the average terrace width is two orders of magnitude smaller than the spot size, we assume that the photoemission signal is also divided equally between both rotational domains. Nevertheless, the finite spin polarization observed here clearly demonstrates that the states at K and K' cannot possess opposite spin textures, since such a configuration would result in complete cancellation of the measured polarization signal.

    The out-of-plane spin polarization $P_z$ extracted from the MDC is presented in Fig. \ref{fig:spin_BME_KMK}\,(g). In contrast to the theoretical expectations, which predict a solely in-plane polarization \cite{Tilgner2025}, a significant out-of-plane spin polarization is observed for the bismuthene valence bands, with a sign reversal along the $\overline{\text{KMK'}}$ direction. Specifically, the bands located along $\overline{\text{KM}}$ exhibit opposite out-of-plane polarization compared to those along $\overline{\text{MK'}}$. The corresponding $P_z$-resolved component intensities are shown in Fig. \ref{fig:spin_BME_KMK}\,(h). Here, $P_z > 0$ corresponds to spin polarization pointing out of the surface plane toward the graphene, whereas $P_z < 0$ indicates polarization directed toward the substrate. At the investigated binding energy, a clear out-of-plane polarization is observed predominantly for the B1 band, while B2 remains nearly unpolarized.

    %Unlike the EDCs discussed previously, where both bands were probed at equal distances from the K point, the MDC intersects B2 substantially closer to K than B1. We therefore hypothesize that the emergence of a finite out-of-plane spin polarization is related to higher-order contributions in $k$ to the effective Hamiltonian, which become increasingly important at larger momenta away from the K point.

    Importantly, the observations presented here are consistent with the out-of-plane spin polarizations extracted from the EDC measurements discussed above (see section IV of the Supplementary Information). In particular, perpendicular to the $\overline{\Gamma\text{K}}$ direction, a finite out-of-plane polarization is observed that is parallel for both bands and reverses sign across the K point. In contrast, no out-of-plane polarization is detected along the $\overline{\Gamma\text{K}}$ direction, consistent with the mirror symmetry of bismuthene.

    However, the observed behavior appears to contradict time-reversal (TR) symmetry. In TR-symmetric systems, the spin textures at $\boldsymbol{k}$ and $-\boldsymbol{k}$ are related by inversion of the complete three-dimensional spin vector, i.e.\ $\boldsymbol{S} \rightarrow -\boldsymbol{S}$ \cite{bernevig2013topological}. While this condition is fulfilled for the measured in-plane spin texture, the observed out-of-plane component does not satisfy this symmetry relation.

    The breaking of TR symmetry indicates final state effects in the photoemission process that generate the finite $P_z$ polarization, rather than representing an initial state property of bismuthene. Previous works on materials with strong SOC \cite{yaji2017spin, heinzmann2012spin, bentmann2017strong, moser2023toy, kuroda2016coherent, dil2019spin} demonstrated that the mutual interference of multiple photoemission channels can generate a final state spin polarization that differs from the initial state. We hypothesize that a similar process is at work in the present case. However, due to the excellent agreement between experiment and theory for the in-plane spin texture, this process appears to affect only the out-of-plane spin component. Future studies may focus on elucidating the presence of final state effects in spin-resolved ARPES on bismuthene by investigating the dependence of the photoelectrons' polarization on light polarization and photon energy. \\

    Finally, the low-energy electronic structure of graphene was investigated by spin-resolved ARPES for both the precursor and bismuthene phases (see section V of the Supplementary Information). In neither case was a significant spin polarization of the graphene \textpi-bands detected.

%=================
%	3 - Conclusion 
%=================	

\section{Summary}

    \noindent
    The spin polarization of the low-energy electronic states of the quantum spin Hall insulator bismuthene and its precursor phase was investigated by means of spin- and angle-resolved photoelectron spectroscopy. For the precursor phase, we demonstrated experimentally and theoretically — for the first time — an energy splitting of the valence band that lifts spin degeneracy. Analysis of the spin orientation revealed a tangential Rashba-type spin splitting.

    For bismuthene, the theoretically predicted in-plane spin polarization of the Dirac-like valence bands \cite{Tilgner2025} was confirmed experimentally, strongly indicating a cubic Rashba-type spin splitting, including higher-order terms in $k$ that lead to deviations from a purely tangential spin texture.

    In addition, we observed a finite out-of-plane spin polarization in bismuthene. This finding is not accounted for by current theoretical considerations of the initial state and appears to be at odds with time-reversal symmetry. We attribute this observation, hypothetically, to the presence of final state interference effects in the photoemission process.

%=================
%	Author contributions
%=================	

\section*{Author contributions}

    \noindent
    N.T., S.G., T.S., and P.S. conceived the project.
    Samples were prepared by N.T.
    Spin-integrated ARPES measurements were carried out by N.T., C.P., F.G., and P.S., while spin-resolved ARPES measurements were performed by N.T., S.W., M.L., T.S., and P.S.
    The experimental data was analyzed by N.T., with significant contributions from S.W., T.S., and P.S.
    DFT calculations were performed and interpreted by A.D.P.U. and S.G.
    All authors discussed the results. N.T. prepared the figures and wrote the manuscript with significant input from F.G. and P.S.

%=================
%	Acknowledgments
%=================	

\section*{Acknowledgments}	

    \noindent
    The authors thank Christoph Lohse for valuable contributions to the substrate preparation. We further acknowledge the MAX IV Laboratory for providing access to the "Bloch" beamline under proposals 20240164, 20250507, and 20261214. In addition, we thank the Bloch beamline staff, Balasubramanian Thiagarajan, and Jacek Osiecki, for their continuous technical and scientific support. This work was funded by the German Research Foundation (Deutsche Forschungsgemeinschaft, DFG) within the framework of Research Unit FOR5242 (project 449119662).

%=================
%	Competing Interests
%=================	

\section*{Declaration of Competing Interest}

    \noindent
    The authors declare that they have no known competing financial interests or personal relationships that could have appeared to influence the work reported in this paper.

%=================
%	Data availability
%=================	

\section*{Data availability}

    \noindent
    Data will be made available on request.

%=================
%	Methods
%=================	

\section*{Methods}

	\subsection*{Sample Preparation}

    \noindent
    4H-SiC wafer pieces were purchased from Pam-Xiamen. Epitaxial graphene was synthesized by polymer-assisted sublimation growth, following the procedure described in Refs. \cite{Kruskopf2DMater.2016, guse2025growth}. Bi intercalation was achieved using an \textit{in situ} deposition and annealing protocol established in earlier studies \cite{SohnJ.KoreanPhys.Soc.2021, Wolff_2024}. Bi deposition was carried out in a dedicated evaporation chamber with a base pressure below $5 \times 10^{-9}$\,mbar using a custom-built Knudsen cell operated at $550^\circ$C for 120\,min. After \textit{in vacuo} transfer to the ultra-high vacuum (UHV) analysis chamber ($2 \times 10^{-10}$\,mbar), Bi intercalation was induced by annealing the sample at $450^\circ$C for 30\,min. The precursor phase was subsequently prepared by an additional annealing step at $950^\circ$C for 10\,min. Sample temperatures were monitored using a pyrometer assuming an emissivity of 0.9. Hydrogen intercalation was performed employing a contactless infrared heating system, in which the sample was annealed at $550^\circ$C for 90\,min under an ultra-pure hydrogen atmosphere at 850\,mbar with a gas flow of 0.9\,slm.

    For spin-resolved ARPES measurements, the samples were transported to the synchrotron facility via a dedicated UHV transport system ($2 \times 10^{-11}$\,mbar). Prior to measurements, bismuthene (precursor) samples were degassed on-site at $200^\circ$C ($400^\circ$C).

    \subsection*{Computational Methods}

    \noindent
    Density functional theory calculations were carried out using the ABINIT package \cite{gonze2009abinit}. The exchange-correlation effects were treated within the generalized gradient approximation (GGA) using the Perdew-Burke-Ernzerhof functional \cite{perdew1996generalized} together with fully relativistic norm-conserving pseudopotentials. Spin-orbit coupling was included in all calculations. Structural relaxations were performed employing an $8 \times 8 \times 1$ Monkhorst-Pack $k$-point mesh until the residual atomic forces were below \qty{5e-6}{\eV\per\angstrom}. The electronic ground-state calculations were subsequently obtained using a denser $12 \times 12 \times 1$ $k$-point grid. A plane-wave cutoff energy of 1088\,eV was adopted throughout the calculations. The spin textures were analyzed using the PyProcar package \cite{pyprocar}. To reduce the computational cost, the graphene overlayer was not included. The dangling bonds at the bottom carbon-terminated surface of the SiC slab were saturated with hydrogen atoms, and a vacuum region of 15\,\AA\ was introduced to avoid interactions between periodic images.

    \subsection*{Spin- and Angle-resolved Photoelectron Spectroscopy}\label{app:SARPES}

    \begin{figure*}[ht!]
        \centering
        \includegraphics[scale=1]{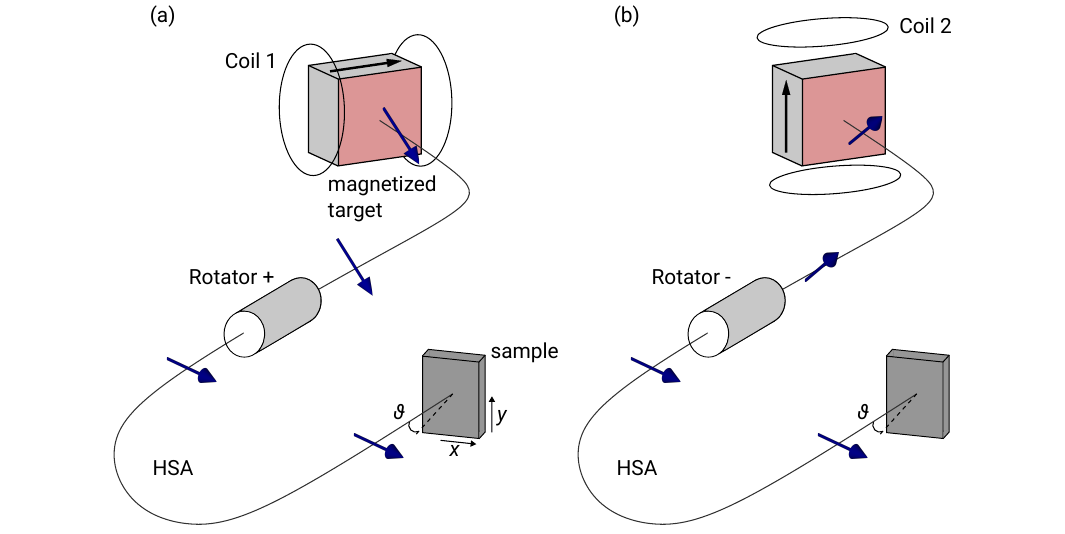}
        \caption{\textbf{Basic setup of the spin-resolved ARPES experiment.}
        Photoelectrons emitted from the sample at a polar angle $\vartheta$ follow the trajectory indicated by the gray solid line. After passing through the hemispherical analyzer (HSA), the magnetic rotator acts on the transverse spin component of the electrons, by rotating it either clockwise (Rotator$+$ (a)) or counterclockwise (Rotator$-$ (b)). The electrons subsequently impinge on a target magnetized along an in-plane direction by means of one of two Helmholtz coil pairs, namely Coil 1 (a) or Coil 2 (b). The black arrow shown on the target denotes the corresponding magnetization direction.
        After Ref. \cite{BlochDocs}.
        }
        \label{fig:SARPES_intro}
    \end{figure*}

    \noindent
    All measurements were carried out at the beamline "Bloch" of the MAX IV Laboratory in Lund, Sweden. Spin-integrated ARPES measurements were performed at the high-resolution A endstation equipped with a DA30-L electron analyzer from Scienta Omicron. Following initial sample alignment, photon energy-dependent measurements in the range of $20\,\text{eV} \leq h\nu \leq 200\,\text{eV}$ and with $p$-polarization were conducted to optimize the photoionization cross-section of the Bi-related electronic states. Energy-momentum maps along the high-symmetry directions were subsequently recorded at selected photon energies.

    Spin-resolved ARPES measurements were performed at the B endstation, which is equipped with a Phoibos 150 hemispherical analyzer from SPECS coupled to both a CCD detector and a Ferrum VLEED spin detector located at the analyzer exit slit. To optimize measurement efficiency, both detectors were operated sequentially. First, spin-integrated cuts through the valence-band structure were acquired as a reference to set up an appropriate measurement geometry. In addition to mechanical sample rotations, an electrostatic deflector was employed to steer the photoelectrons, thereby effectively tuning the in-plane momentum corresponding to polar and tilt rotations. The combination of manipulator motions and deflector settings enabled precise selection of the momentum positions for the subsequent acquisition of spin-resolved EDCs.

    The VLEED spin detector exploits the spin-dependent reflectivity of electrons from a magnetized target. The resulting asymmetry in the scattering probability originates from the unequal density of states for opposite spin directions in the ferromagnetic target material \cite{okuda2015double, okuda2013spin}. As target, a thin oxidized Fe film grown on a W(001) crystal was used. The target could be magnetized along two orthogonal in-plane directions by means of two Helmholtz coil pairs, denoted Coil 1 (C1) and Coil 2 (C2). The scattered electron intensity was detected using a channeltron.

    In addition to the target magnetization, a magnetic spin rotator -- located between the hemispherical analyzer and the spin detector -- acted on the transverse spin component of the electron beam. Depending on the operation mode, the spin was rotated either clockwise (Rotator$+$ (R$+$)) or counterclockwise (Rotator$-$ (R$-$)). Fig. \ref{fig:SARPES_intro} illustrates two representative combinations of coil and rotator settings: C1R$+$ in (a) and C2R$-$ in (b). By combining all four configurations (C1R$+$, C1R$-$, C2R$+$, and C2R$-$), the complete 3D spin polarization vector of the photoelectron beam could be determined, as outlined below.

    For each measurement configuration, the spin asymmetry $A$ was calculated from the intensities measured for positive ($I_+$) and negative ($I_-$) target magnetization according to
    \begin{equation*}
        A = \frac{I_+ - I_-}{I_+ + I_-}.
    \end{equation*}

    The in-plane spin polarization components $P'_x$ and $P'_y$ in the detector coordinate system were determined from the difference and sum, respectively, of the asymmetries measured using C2 with R$+$ and R$-$:
    \begin{equation*}
        P'_x = -\frac{A_{\text{C2R}+} - A_{\text{C2R}-}}{\sqrt{2} S}
        \,\, \text{and} \,\,
        P'_y = \frac{A_{\text{C2R}+} + A_{\text{C2R}-}}{\sqrt{2} S}.
    \end{equation*}
    Here, $S = 0.29$ denotes the Sherman function, which quantifies the analyzing power of the spin detector \cite{jakubassa2012sherman}. It corresponds to the asymmetry expected for a fully spin-polarized electron beam \cite{dil2009spin}.

    The out-of-plane component $P'_z$ was obtained analogously using the C1 configuration. In this case, an additional geometrical correction accounting for the finite angle $\delta = 15^\circ$ between the incoming electron beam and the surface normal of the target had to be included:
    \begin{equation*}
        P'_z = \frac{1}{\cos{\delta}} \left( \frac{A_{\text{C1R}+} + A_{\text{C1R}-}}{2S}
                - \frac{\sin{\delta}}{\sqrt{2}} P'_x \right).
    \end{equation*}

    Up to this point, the discussion refers to the polarization vector in the detector coordinate system, $\boldsymbol{P}' = (P'_x,P'_y,P'_z)^\text{T}$. Experimentally, however, the relevant quantity is the polarization vector in the sample reference frame, $\boldsymbol{P} = (P_x,P_y,P_z)^\text{T}$. For measurements in normal emission, both coordinate systems coincide. For a finite polar sample rotation $\vartheta$, the measured polarization vector $\boldsymbol{P}'$ must be transformed by a rotation around the $y$-axis (see Fig. \ref{fig:SARPES_intro}\,(a)), which is the same in both coordinate systems. The polarization vector in the sample frame is therefore given by
    \begin{equation*}
        \begin{pmatrix}
            P_x \\
            P_y \\
            P_z
        \end{pmatrix}
        = 
        \begin{pmatrix}
            \cos{\vartheta} & 0 & \sin{\vartheta} \\
            0 & 1 & 0 \\
            -\sin{\vartheta} & 0 & \cos{\vartheta}
        \end{pmatrix}
        \cdot
        \begin{pmatrix}
            P'_x \\
            P'_y \\
            P'_z
        \end{pmatrix}.
    \end{equation*}

    This correction is not required for momentum shifts introduced by the electrostatic deflector, since the steering process does not alter the spin orientation of the electrons.

    Finally, the total spin-integrated intensity $I_\text{tot}$ can be decomposed into contributions corresponding to positive and negative spin polarization according to
    \begin{equation*}
        C_{\pm i} = \frac{I_\text{tot}}{2} (1 \pm P_i)
        \,\,\text{with}\,\,
        i \in \{x,y,z\}.
    \end{equation*}

    The samples were kept at a constant temperature of 20\,K during all experiments.

%=================
%	Literature
%=================	

%apsrev4-2.bst 2019-01-14 (MD) hand-edited version of apsrev4-1.bst
%Control: key (0)
%Control: author (8) initials jnrlst
%Control: editor formatted (1) identically to author
%Control: production of article title (0) allowed
%Control: page (0) single
%Control: year (1) truncated
%Control: production of eprint (0) enabled
%

\end{document}

\title{Supplementary Information for: \\
Spin- and Angle-resolved Photoelectron Spectroscopy Study of the Quantum Spin Hall Insulator Bismuthene and its Precursor}

\author{Niclas Tilgner}
\email{niclas.tilgner@physik.tu-chemnitz.de}
\affiliation{Institute of Physics, Chemnitz University of Technology, 09126 Chemnitz, Germany}
\affiliation{Research Center for Materials, Architectures and Integration of Nanomembranes (MAIN), 09126 Chemnitz, Germany}

\author{Andres David Pe\~{n}a Unigarro}
\affiliation{Institute of Physics, Chemnitz University of Technology, 09126 Chemnitz, Germany}

\author{Susanne Wolff}
\affiliation{Institute of Physics, Chemnitz University of Technology, 09126 Chemnitz, Germany}
\affiliation{Research Center for Materials, Architectures and Integration of Nanomembranes (MAIN), 09126 Chemnitz, Germany}

\author{Mats Leandersson}
\affiliation{MAX IV Laboratory, Lund University, 22484 Lund, Sweden}

\author{Craig Polley}
\affiliation{MAX IV Laboratory, Lund University, 22484 Lund, Sweden}

\author{Fabian G\"{o}hler}
\affiliation{Institute of Physics, Chemnitz University of Technology, 09126 Chemnitz, Germany}
\affiliation{Research Center for Materials, Architectures and Integration of Nanomembranes (MAIN), 09126 Chemnitz, Germany}

\author{Sibylle Gemming}
\affiliation{Institute of Physics, Chemnitz University of Technology, 09126 Chemnitz, Germany}

\author{Thomas Seyller}
\affiliation{Institute of Physics, Chemnitz University of Technology, 09126 Chemnitz, Germany}
\affiliation{Research Center for Materials, Architectures and Integration of Nanomembranes (MAIN), 09126 Chemnitz, Germany}

\author{Philip Sch\"{a}dlich}
\email{philip.schaedlich@physik.tu-chemnitz.de}
\affiliation{Institute of Physics, Chemnitz University of Technology, 09126 Chemnitz, Germany}
\affiliation{Research Center for Materials, Architectures and Integration of Nanomembranes (MAIN), 09126 Chemnitz, Germany}

\date{\today}

\maketitle

\section{DFT of the precursor phase}

\begin{figure*}[ht!]
        \centering
        \includegraphics[scale=1]{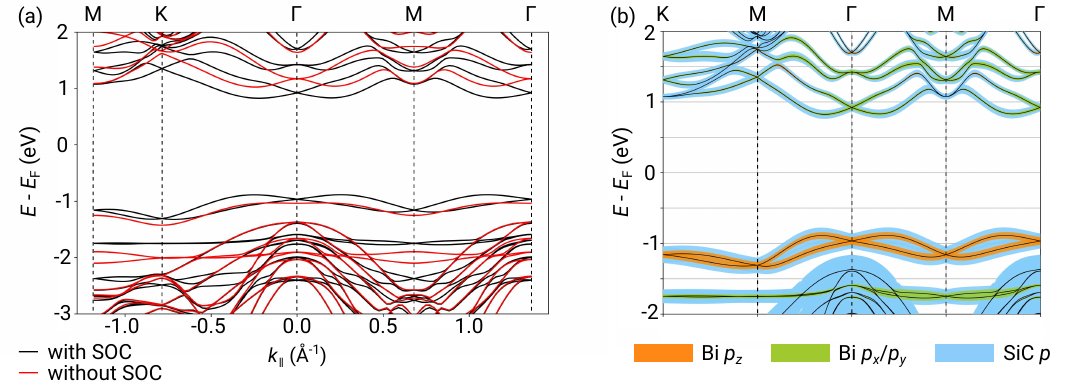}
        \caption{\textbf{DFT of the precursor phase.}
        (a) Calculated low-energy electronic structure of the precursor phase along the high-symmetry directions of the Bi lattice with (black) and without (red) considering spin-orbit coupling (SOC), respectively. The graphene was omitted in the calculation.
        (b) Orbital projection of Bi $p_z$- (orange), Bi $p_x$- and $p_y$- (green), and SiC $p$-orbitals (blue) onto the calculated band structure. The width of the colored bands is proportional to the weight of the respective orbital contribution.
        Vertical, black dashed lines indicate high-symmetry points of the Bi adatom structure.
        }
        \label{fig:DFT_PRC}
    \end{figure*}

    \noindent
    The low-energy electronic structure of the precursor phase was calculated using DFT with the structural model depicted in Fig. 1\,(b) of the main text. This model assumes a 1/3 monolayer of Bi atoms on T$_4$ sites of the 4H-SiC(0001) surface. Graphene was omitted from the calculation. The obtained band structure is shown in Fig. \ref{fig:DFT_PRC}\,(a), with and without considering SOC, in black and red lines, respectively. Note that to match the experimentally observed dispersion, this calculation was offset by 0.3\,eV in Fig. 1\,(a) of the main text. We hypothesize that this offset is related to the graphene, which was neglected in the calculation and induces a change in the chemical potential at the surface.

    Fig. \ref{fig:DFT_PRC}\,(b) shows the orbital projection of the calculated band structure onto Bi and SiC $p$-orbitals. As described in the main text, Bi adatoms are $sp^3$-hybridized in this configuration, saturating three dangling bonds of the substrate surface. Thus, the in-plane $p_x$- and $p_y$-contributions (green) move away from the Fermi level and hybridize with the SiC $p$-orbitals (blue). The remaining Bi hybrid orbital is fully occupied and points out of the surface plane, giving rise to the observed surface state. Consistent with this, the orbital projection onto the surface state is solely of Bi $p_z$-type (orange).

\newpage
\section{Out-of-plane spin polarization of the precursor phase}

\begin{figure*}[ht!]
        \centering
        \includegraphics[scale=1]{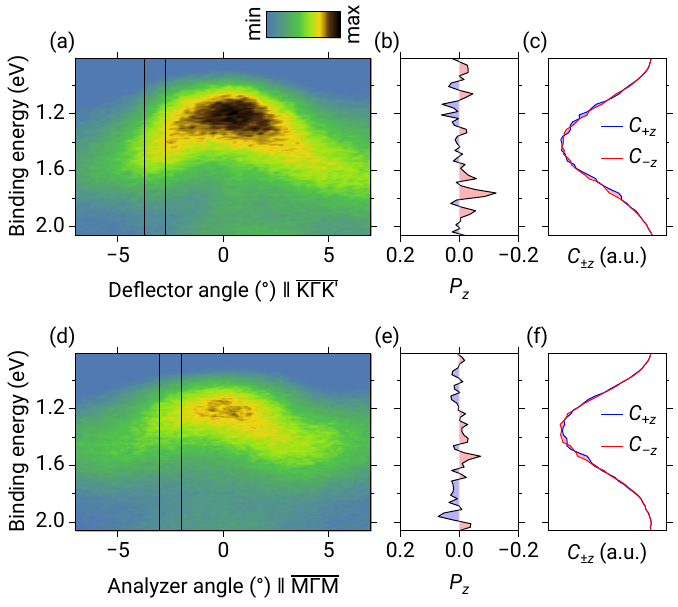}
        \caption{\textbf{Out-of-plane spin polarization of the precursor phase.}
        (a,d) Energy-angle maps measured along the $\overline{\text{K}\Gamma\text{K'}}$ direction (a) and the $\overline{\text{M}\Gamma\text{M}}$ direction (d). The vertical black lines indicate the regions from which the EDCs were extracted with an angular integration width of $1^\circ$.
        (b-c) and (e-f) Corresponding out-of-plane spin polarizations $P_z$ together with the derived spin-resolved component intensities $C_{\pm z}$. Blue (red) represents positive (negative) spin polarization.
        All measurements were performed using a photon energy of $h\nu = 152.0$\,eV and $p$-polarization.
        }
        \label{fig:spin_PRC_Pz}
    \end{figure*}

    \noindent
    The main text focuses on the in-plane spin polarizations of the two EDCs acquired for the precursor phase. Here, in Figs. \ref{fig:spin_PRC_Pz}\,(b) and (e), we additionally present the out-of-plane spin polarization ($P_z$) measured at the same momentum positions and extracted from the same dataset as the in-plane polarizations discussed in the main text. No significant out-of-plane polarization is observed. Consequently, the $P_z$-resolved component intensities $C_{\pm z}$ shown in Figs. \ref{fig:spin_PRC_Pz}\,(c) and (f) do not permit a distinction between states E1 and E2. The spin polarization of the precursor valence bands at the investigated momentum positions is therefore found to be entirely in-plane within the experimental resolution.

\newpage
\section{Complementary Spin-resolved EDC of the bismuthene phase}

\begin{figure*}[ht!]
        \centering
        \includegraphics[scale=1]{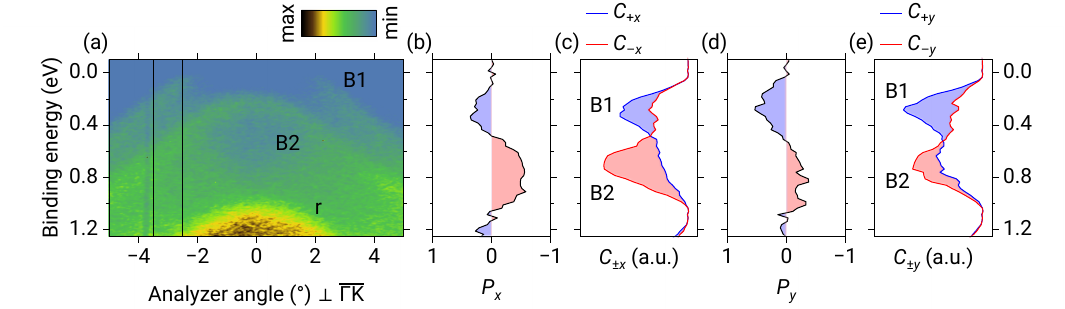}
        \caption{\textbf{In-plane spin polarization of the bismuthene phase.}
        (a) Energy-angle map recorded perpendicular to the $\overline{\Gamma\text{K}}$ direction of bismuthene. The vertical black lines mark the momentum region from which the EDC was extracted.
        (b-e) Corresponding in-plane spin polarizations $P_i$ together with the derived spin-resolved component intensities $C_{\pm i}$ with $i \in \{x,y\}$. Blue (red) denotes positive (negative) spin polarization.
        The photon energy was $h\nu = 39.0$\,eV and the beam was $p$-polarized.
        }
        \label{fig:spin_BME_K_other_side}
    \end{figure*}

    \noindent
    Fig. \ref{fig:spin_BME_K_other_side}\,(a) shows the energy-angle map of bismuthene recorded perpendicular to the $\overline{\Gamma\text{K}}$ direction. In Fig. 4 of the main text only a spin-resolved EDC extracted at positive analyzer angles of this cut is presented. Figs. \ref{fig:spin_BME_K_other_side}\,(b-e) show the analogous measurement on the other side, as depicted by the vertical black lines in (a). Two oppositely polarized contributions shifted in energy are present and assigned to the bismuthene valence bands B1 and B2. The results of Fig. \ref{fig:spin_BME_K_other_side} support the direction of the bottom arrows in Figs. 4\,(a) and (b) of the main text.

\newpage
\section{Out-of-plane spin polarization of the bismuthene phase}

\begin{figure*}[ht!]
        \centering
        \includegraphics[scale=1]{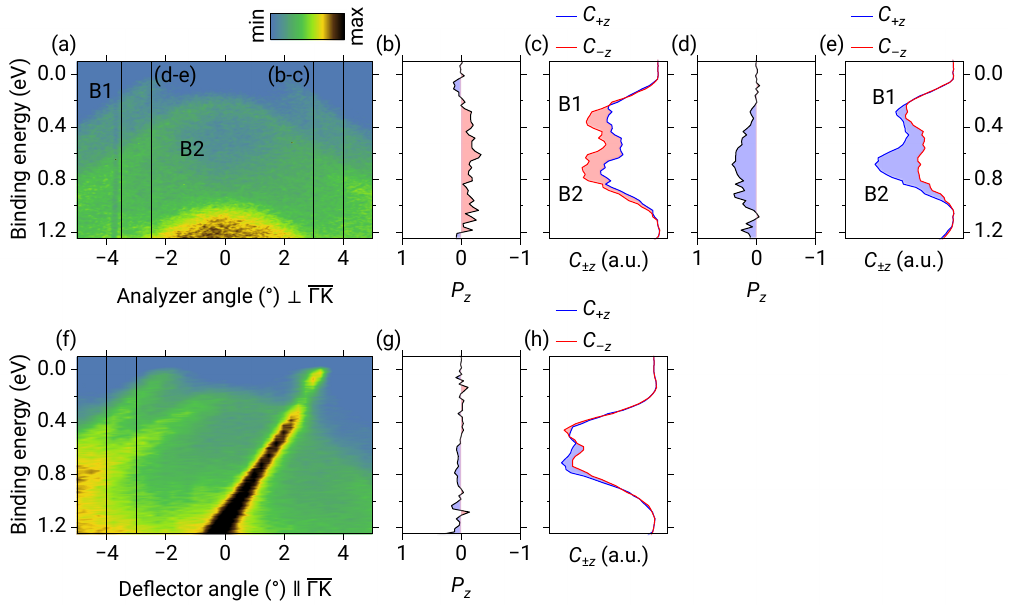}
        \caption{\textbf{Out-of-plane spin polarization of the bismuthene phase.}
        (a,f) Energy-angle maps measured perpendicular (a) and parallel (f) to the $\overline{\Gamma\text{K}}$ direction. Vertical black lines denote the momentum regions from which the EDCs were extracted using an angular integration width of $1^\circ$.
        (b-e) and (g-h) Corresponding out-of-plane spin polarizations $P_z$ together with the derived spin-resolved component intensities $C_{\pm z}$. Blue (red) denotes positive (negative) spin polarization.
        The measurements were performed with a photon energy of $h\nu = 39.0$\,eV and $p$-polarization.
        }
        \label{fig:spin_BME_Pz}
    \end{figure*}

    \noindent
    In Fig. 4 of the main text as well as Fig. \ref{fig:spin_BME_K_other_side}, we present the in-plane spin polarizations of the bismuthene valence bands at three distinct momentum positions. Here, in Figs. \ref{fig:spin_BME_Pz}\,(b), (e), and (g), we additionally show the corresponding out-of-plane spin polarization ($P_z$), measured at the same momentum positions and extracted from the same datasets as the in-plane polarizations.

    Perpendicular to the $\overline{\Gamma\text{K}}$ direction, negative values of $P_z$ are observed for positive analyzer angles (see Fig. \ref{fig:spin_BME_Pz}\,(b)), while positive values of $P_z$ are detected on the opposite side of the K point (see Fig. \ref{fig:spin_BME_Pz}\,(e)) within the energy range of the bismuthene valence bands (compare Fig. \ref{fig:spin_BME_Pz}\,(a)). From the $P_z$-resolved component intensities $C_{\pm z}$ shown in Figs. \ref{fig:spin_BME_Pz}\,(c) and (f), it becomes clear that both valence bands exhibit the same out-of-plane polarization at a given momentum position, whereas the sign of $P_z$ reverses across the K point. This observation is consistent with the spin-resolved MDC measurements discussed in the main text.

    The spin-resolved EDC acquired along $\overline{\Gamma\text{K}}$ (compare Fig. \ref{fig:spin_BME_Pz}\,(f)) does not exhibit a significant out-of-plane polarization (Fig. \ref{fig:spin_BME_Pz}\,(g)). This agrees well with $P_z$ changing sign when moving along the constant-energy contour around the K point and is consistent with the mirror symmetry of the system.

\newpage
\section{Spin-resolved EDCs on graphene}

\begin{figure*}[ht!]
        \centering
        \includegraphics[scale=1]{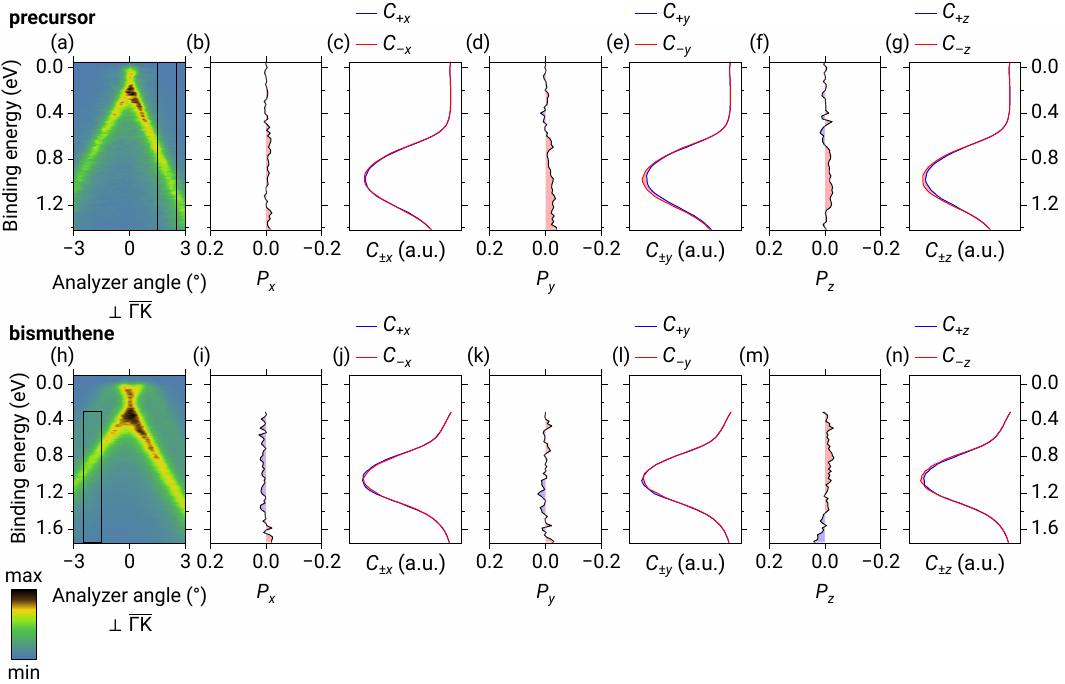}
        \caption{\textbf{Spin-resolved EDCs on graphene.}
        (a,h) Energy-angle maps measured perpendicular to the $\overline{\Gamma\text{K}_\text{Gr}}$ direction of graphene on the precursor (a) and bismuthene (h) phases. Black rectangles denote the momentum regions from which the EDCs were extracted using an angular integration width of $1^\circ$.
        (b-g) and (i-n) Corresponding spin polarizations $P_i$ together with the derived spin-resolved component intensities $C_{\pm i}$ with $i \in \{x,y,z\}$. Blue (red) denotes positive (negative) spin polarization.
        The measurements were performed with a photon energy of $h\nu = 40.0$\,eV and $p$-polarization.
        }
        \label{fig:spin_Gr}
    \end{figure*}

    \noindent
    We additionally investigated a hypothetical spin polarization in the \textpi-bands of the graphene on top of both the precursor and bismuthene phases. To this end, spin-resolved EDCs were acquired at momentum positions in the vicinity of the graphene K$_{\text{Gr}}$ point. Spin-integrated energy-angle maps of the graphene Dirac cones are shown in Figs. \ref{fig:spin_Gr}\,(a) and (h), where the black rectangles indicate the regions from which the spin-resolved EDCs were extracted. Note that the photoemission signal of the n-type Dirac cone of graphene on bismuthene is superimposed with a weaker signal from the p-type Dirac cone of solely H-intercalated graphene. However, for both phases and for all three measured polarization directions, no significant lifting of the \textpi-bands' spin degeneracy was detected (see Figs. \ref{fig:spin_Gr}\,(b-g) and (i-n)).

\newpage
\section{Spin-resolved ARPES Data}

    \noindent
    This section presents the raw spin-resolved ARPES data, i.e.\ the EDCs and MDCs recorded for the different coil and rotator configurations that form the basis of the discussed analysis, in particular Figs. 2, 4, and 5 of the main text as well as Figs. \ref{fig:spin_PRC_Pz} -- \ref{fig:spin_Gr}. For each momentum position in the EDC measurements, or for each selected energy in the MDC measurements, all datasets used for the determination of the spin polarization are displayed in the figures below.

    The figures are arranged in a layout consisting of three columns and four rows. Each row corresponds to one specific combination of rotator setting (Rotator$+$ (R$+$) or Rotator$-$ (R$-$)) and Helmholtz coil configuration (Coil 1 (C1) or Coil 2 (C2)). For each of these four configurations, measurements were carried out using both positive and negative target magnetization (see Methods section of the main text). These two cases are represented by blue (Coil$+$) and orange curves (Coil$-$), respectively.

\begin{figure}[t!]
        \centering
        \includegraphics[width=\textwidth]{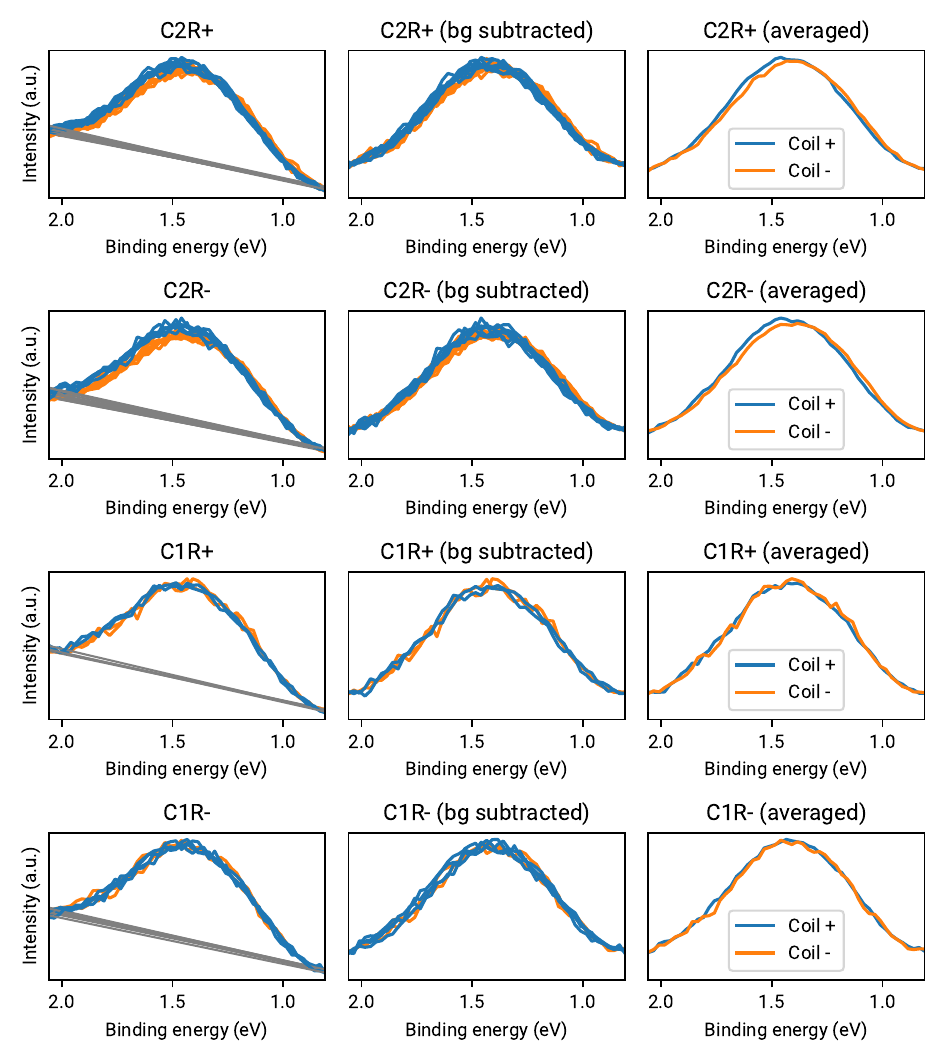}
        \caption{EDCs of the precursor with both coil and rotator settings supporting the plots of Figs. 2\,(d-g) of the main text and Figs. \ref{fig:spin_PRC_Pz}\,(b-c).
        }
        \label{app:spin_PRC_02}
    \end{figure}

\begin{figure}[t!]
        \centering
        \includegraphics[width=\textwidth]{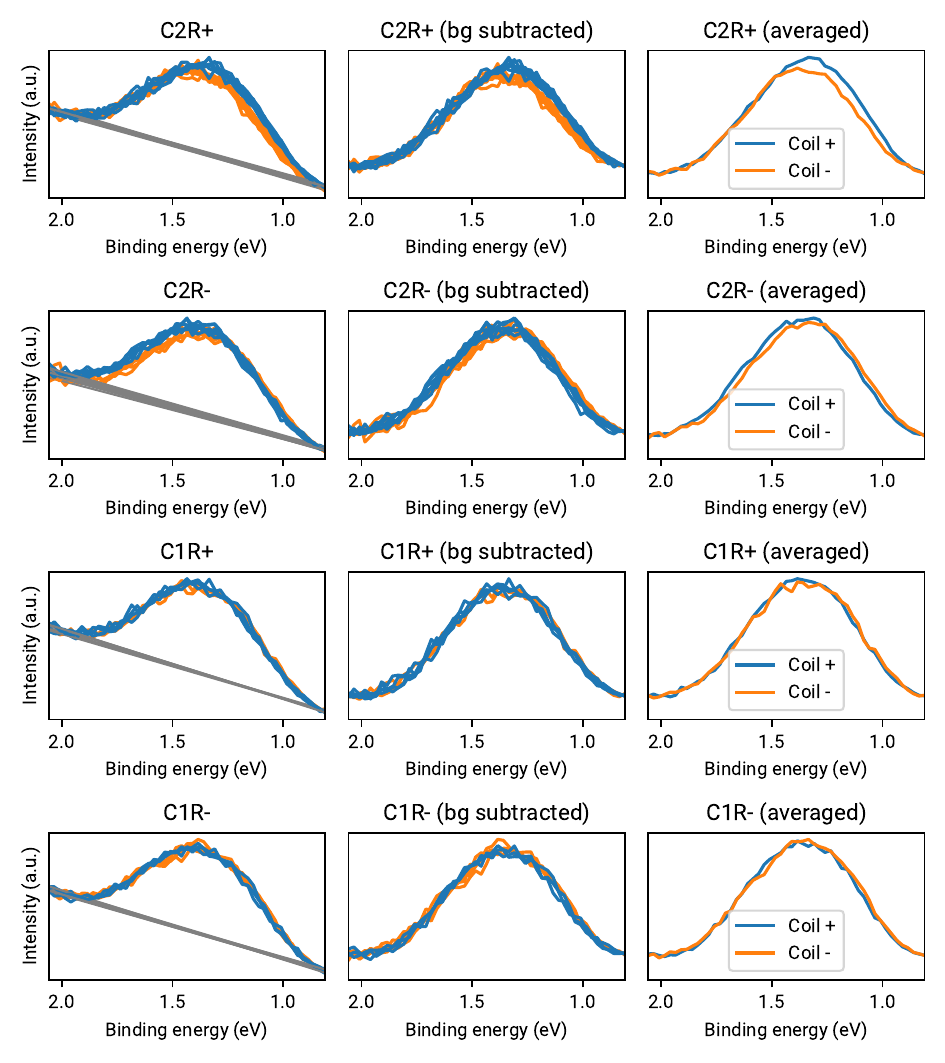}
        \caption{EDCs of the precursor with both coil and rotator settings supporting the plots of Figs. 2\,(i-l) of the main text and Figs. \ref{fig:spin_PRC_Pz}\,(e-f).
        }
        \label{app:spin_PRC_01}
    \end{figure}

\begin{figure}[t!]
        \centering
        \includegraphics[width=\textwidth]{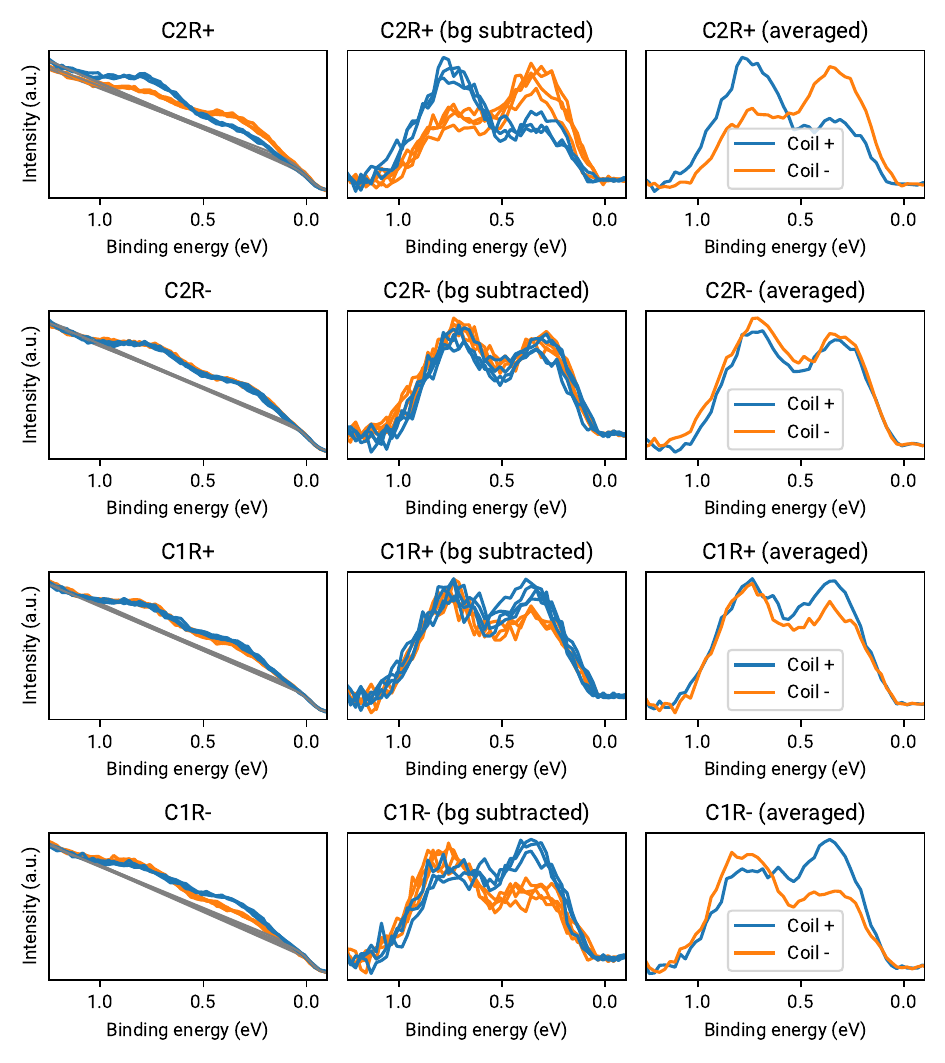}
        \caption{EDCs of bismuthene with both coil and rotator settings supporting the plots of Figs. 4\,(d-g) of the main text and Figs. \ref{fig:spin_BME_Pz}\,(b-c).
        }
        \label{app:spin_BME_01}
    \end{figure}

\begin{figure}[t!]
        \centering
        \includegraphics[width=\textwidth]{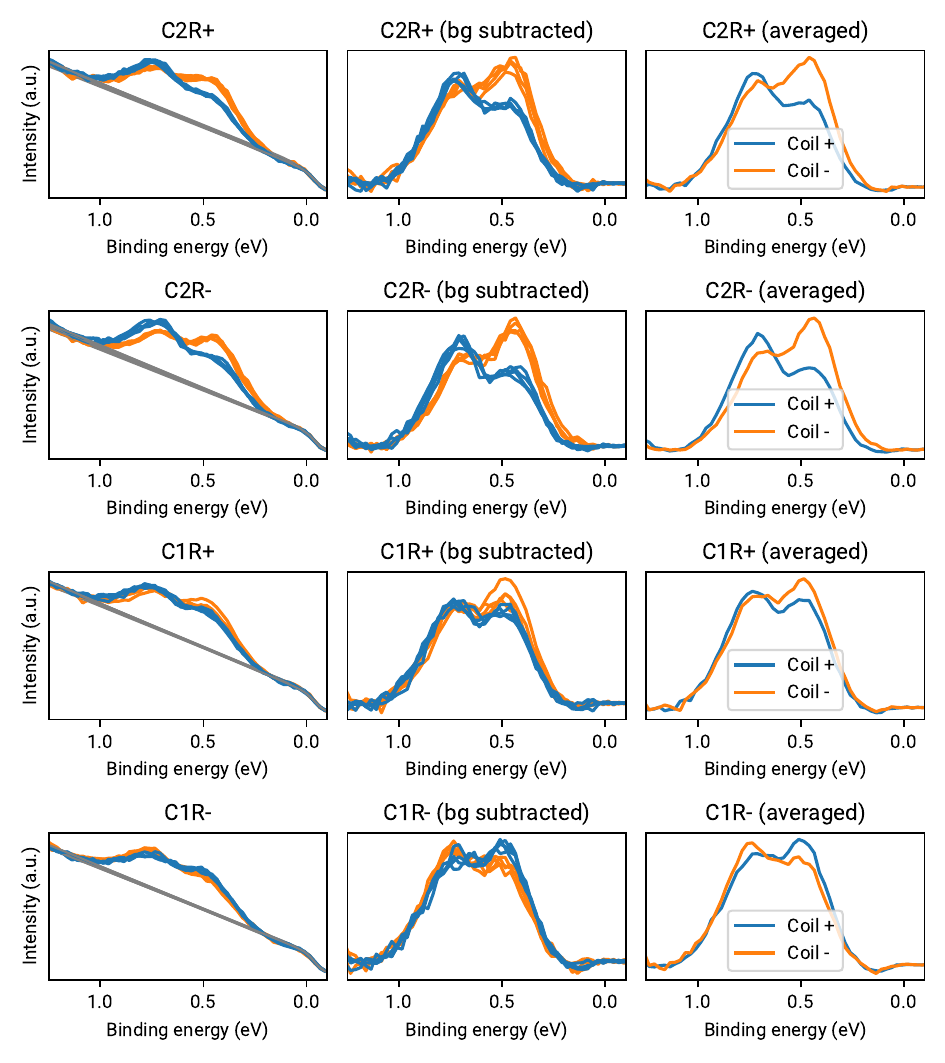}
        \caption{EDCs of bismuthene with both coil and rotator settings supporting the plots of Figs. 4\,(i-l) of the main text and Figs. \ref{fig:spin_BME_Pz}\,(g-h).
        }
        \label{app:spin_BME_02}
    \end{figure}

\begin{figure}[t!]
        \centering
        \includegraphics[width=\textwidth]{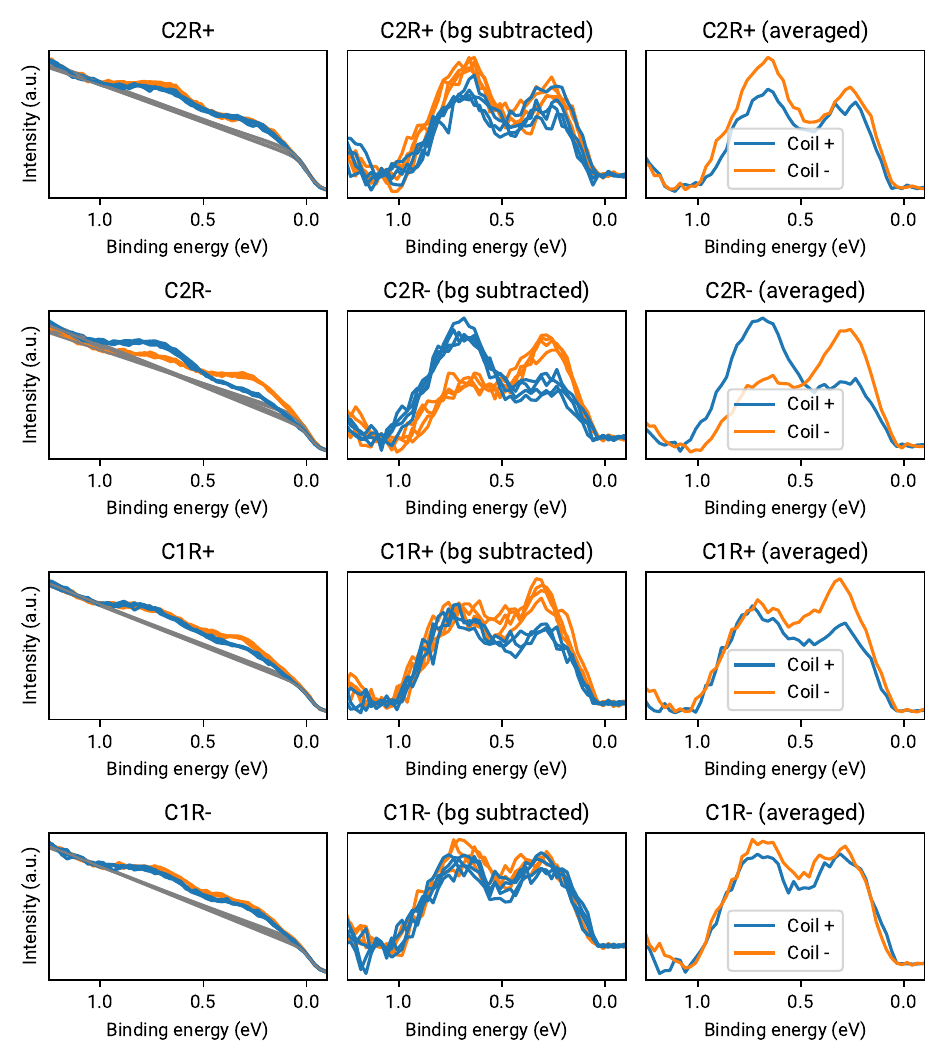}
        \caption{EDCs of bismuthene with both coil and rotator settings supporting the plots of Figs. \ref{fig:spin_BME_K_other_side}\,(b-e) and Figs. \ref{fig:spin_BME_Pz}\,(d-e).
        }
        \label{app:spin_BME_04}
    \end{figure}

\begin{figure}[t!]
        \centering
        \includegraphics[width=\textwidth]{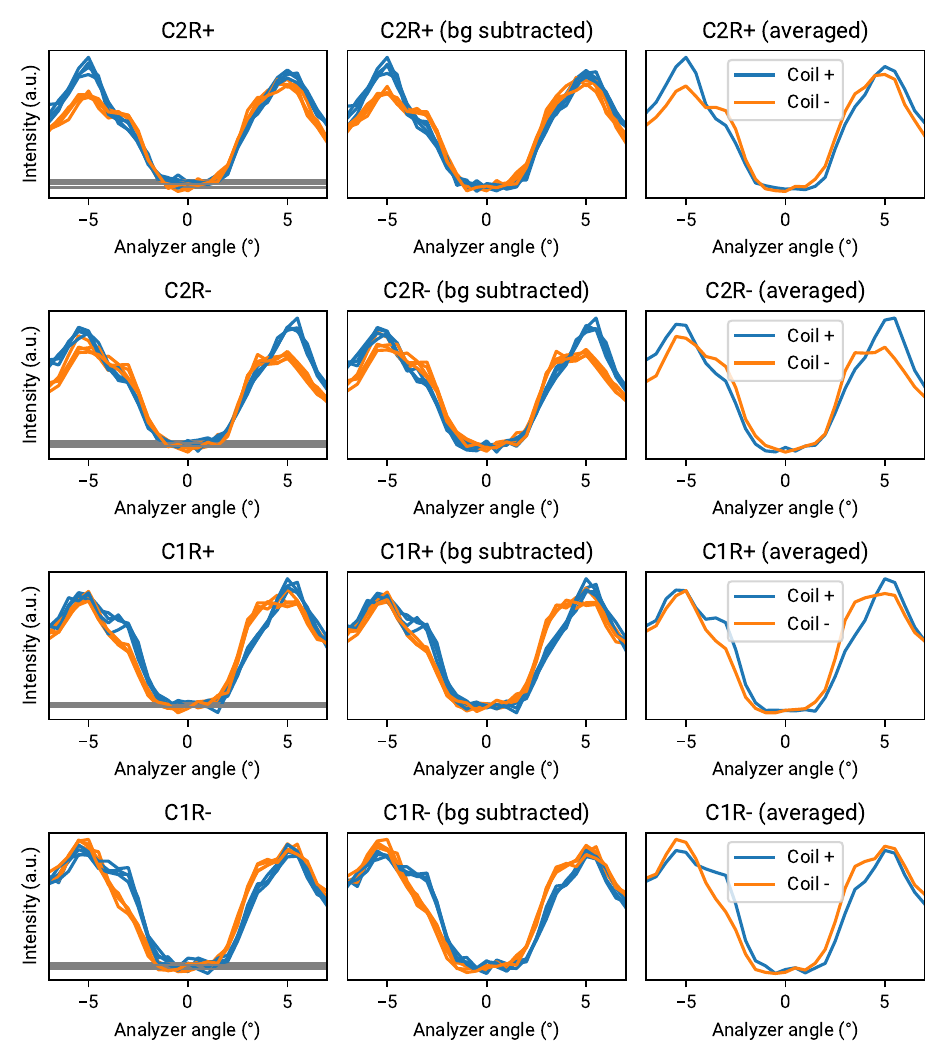}
        \caption{MDCs of bismuthene with both coil and rotator settings supporting the plots of Fig. 5 of the main text.
        }
        \label{app:spin_BME_03}
    \end{figure}

\begin{figure}[t!]
        \centering
        \includegraphics[width=\textwidth]{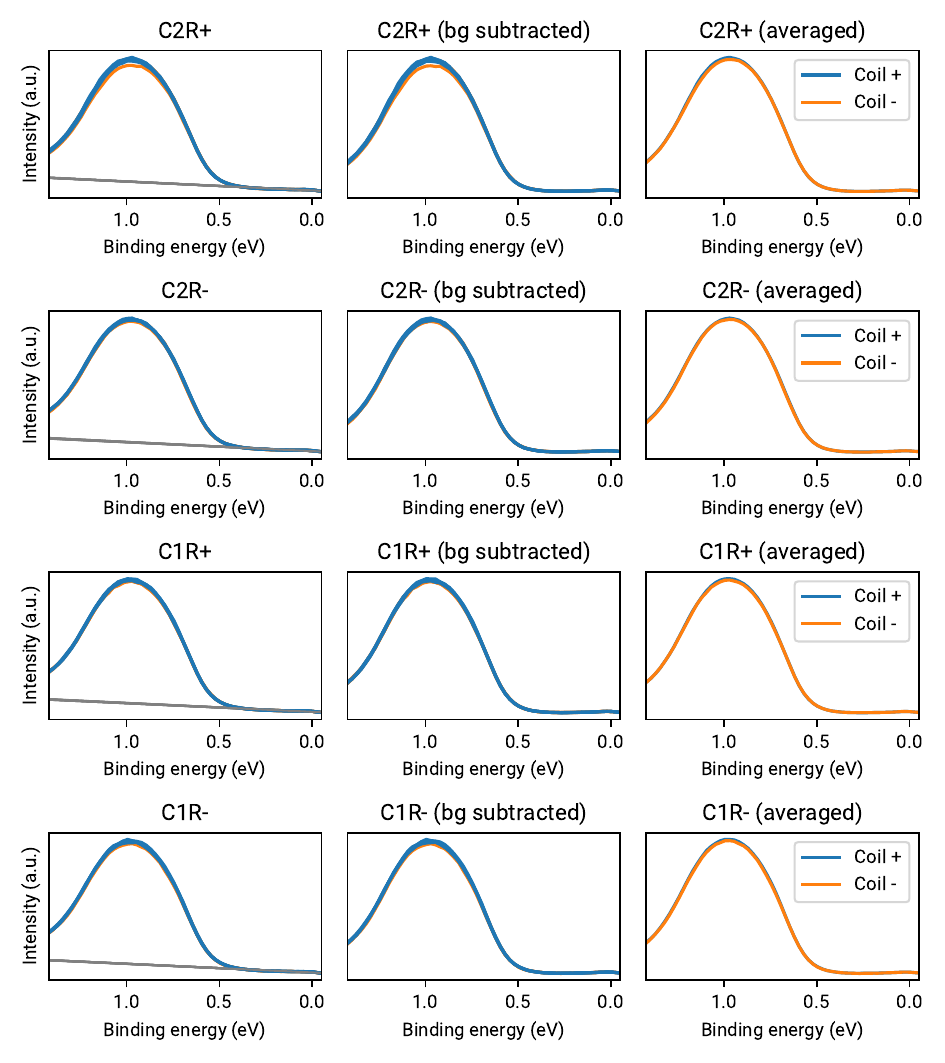}
        \caption{EDCs of graphene on the precursor phase with both coil and rotator settings supporting the plots of Figs. \ref{fig:spin_Gr}\,(b-g).
        }
        \label{app:spin_PRC_Gr}
    \end{figure}

\begin{figure}[t!]
        \centering
        \includegraphics[width=\textwidth]{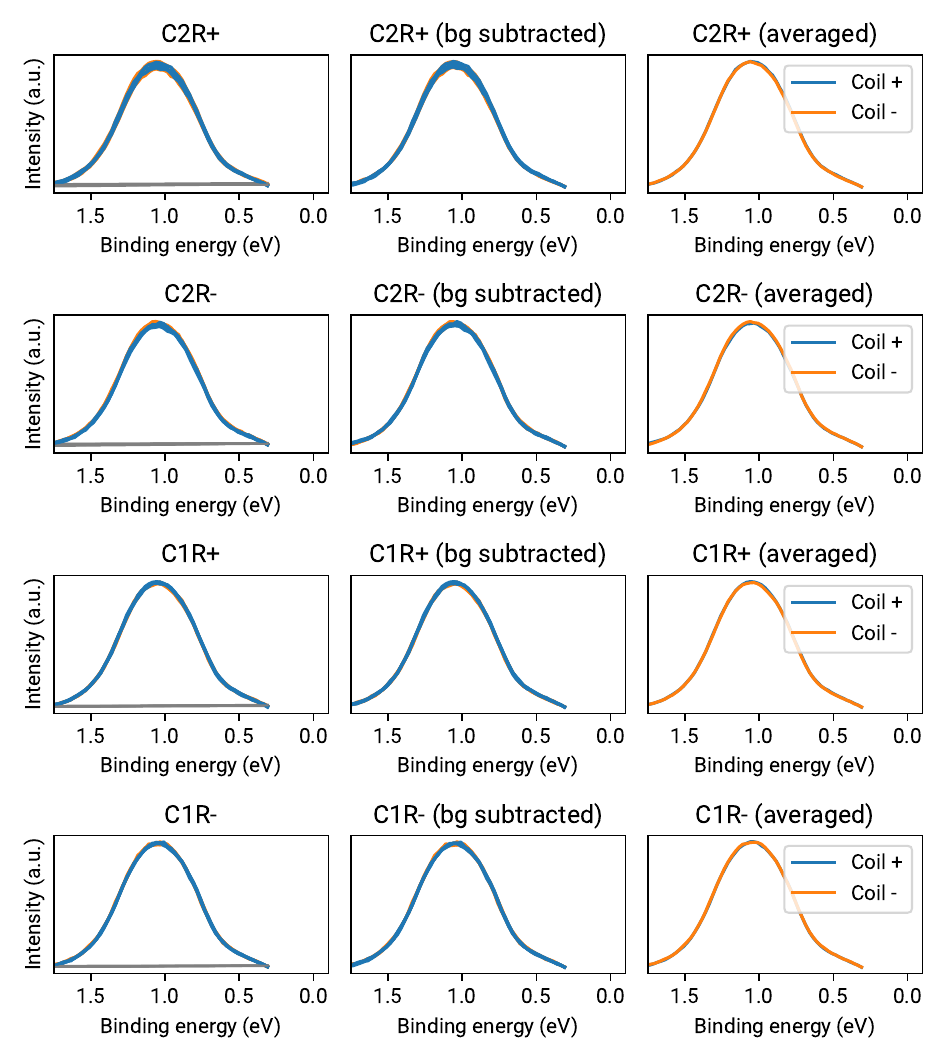}
        \caption{EDCs of graphene on bismuthene with both coil and rotator settings supporting the plots of Figs. \ref{fig:spin_Gr}\,(i-n).
        }
        \label{app:spin_BME_Gr}
    \end{figure}

    Consequently, each EDC or MDC comprises eight distinct measurement configurations. Multiple scans were recorded for every configuration. The individual scans are shown in the first column of the figures below. For each scan, a background was determined in order to separate the inelastic contribution from the photoemission signal. The background functions are overlaid as gray curves in the first column. For the two precursor phase as well as the two graphene EDC measurements (see Figs. \ref{app:spin_PRC_02}, \ref{app:spin_PRC_01}, \ref{app:spin_PRC_Gr}, and \ref{app:spin_BME_Gr}), a purely linear background was employed. In contrast, the EDCs measured on bismuthene (see Figs. \ref{app:spin_BME_01} -- \ref{app:spin_BME_04}) were fitted using a linear function multiplied by a Fermi-Dirac distribution in order to reproduce the background shape more accurately. For the bismuthene MDC measurement (see Fig. \ref{app:spin_BME_03}), a constant background was used.

    The second column of each figure ("bg subtracted") displays the individual scans after subtraction of the corresponding background contribution. Averaging all scans recorded under identical measurement conditions yields the spectra shown in the third column ("averaged"). These averaged curves were subsequently used to calculate the spin polarizations according to the formalism described in the Methods section of the main text.

